\documentclass{optica-article}

\journal{opticajournal} 

\articletype{Research Article}

\usepackage{lineno}

\usepackage[final]{pdfpages}

\usepackage{fontawesome}
\usepackage{float}
\usepackage{amsmath}
\usepackage{bm}

\usepackage{acronym}		
\usepackage{glossaries}		
\usepackage{booktabs}

\newcommand{\fib}{fibre} 

\newacronym{USyd}{USyd}{the University of Sydney}
\newacronym{ANU}{ANU}{Australia National University}
\newacronym{ESO}{ESO}{European Southern Observatory}

\newacronym{NSF}{NSF}{National Science Foundation}
\newacronym{LIEF}{LIEF}{Linkage Infrastructure, Equipment and Facilities}

\newacronym{VLTI}{VLTI}{Very Large Telescope Interferometer}
\newacronym{ATs}{ATs}{auxiliary telescopes}
\newacronym{UTs}{UTs}{unit telescopes}
\newacronym{ELT}{ELT}{Extremely Large Telescope}
\newacronym{JWST}{JWST}{James Webb Space Telescope}
\newacronym{DCT}{LDT}{Lowell Discovery Telescope}
\newacronym{GMT}{GMT}{Giant Magellan Telescope}
\newacronym{LBT}{LBT}{Large Binocular Telescope}

\newacronym{STS}{STS}{six telescope simulator}
\newacronym{GPAO}{GPAO}{GRAVITY+ adaptive optics}

\newacronym{WFS}{WFS}{wavefront sensor}
\newacronym{ADC}{ADC}{atmospheric dispersion corrector}
\newacronym{LDC}{LDC}{longitudinal dispersion corrector}
\newacronym{AO}{AO}{adaptive optics}
\newacronym{SNR}{SNR}{signal to noise ratio}
\newacronym{ADU}{ADU}{arbitrary data units}
\newacronym{RMS}{RMS}{root mean square}

\newacronym{OAP}{OAP}{off-axis paraboloid}
\newacronym{DM}{DM}{deformable mirror}
\newacronym{MEMS}{MEMS}{micro-electromechanical system}
\newacronym{IR}{IR}{infrared}
\newacronym{SLM}{SLM}{spatial light modulator}

\newacronym{SMF}{SMF}{single-mode \fib{}}
\newacronym{MCF}{MCF}{multicore \fib{}}
\newacronym{MMF}{MMF}{multimode \fib{}}
\newacronym{SLD}{SLD}{superluminescent diode}

\newacronym{CAD}{CAD}{computer aided design}
\newacronym{GUI}{GUI}{graphical user interface}

\begin{document}

\title{Illuminating the lantern: coherent, spectro-polarimetric characterisation of a multimode converter}

\author{Adam K. Taras,\authormark{1,2,*} Barnaby R. M. Norris,\authormark{1,2} Christopher Betters,\authormark{1,2} Andrew Ross-Adams,\authormark{1,2} Peter G. Tuthill,\authormark{1,2} Jin Wei,\authormark{1,2} and Sergio Leon-Saval\authormark{1,2}}

\address{\authormark{1}Sydney Institute for Astronomy, School of Physics, The University of Sydney, NSW 2006, Australia\\
\authormark{2}Sydney Astrophotonics Instrumentation Laboratory, School of Physics, The University of Sydney, NSW 2006, Australia}

\email{adam.taras@sydney.edu.au}


\begin{abstract*} 
While photonic lanterns efficiently and uniquely map a set of input modes to single-mode outputs (or vice versa), the optical mode transfer matrix of any particular fabricated device cannot be constrained at the design stage due to manufacturing imperfections. Accurate knowledge of the mapping enables complex sensing or beam control applications that leverage multimode conversion. In this work, we present a characterisation system to directly measure the electric field from a photonic lantern using digital off-axis holography, following its evolution over a 73\,nm range near 1550\,nm and in two orthogonal, linear polarisations. We provide the first multi-wavelength, polarisation decomposed characterisation of the principal modes of a photonic lantern. Performance of our testbed is validated on a single-mode fibre then harnessed to characterise a 19-port, multicore fibre fed photonic lantern. We uncover the typical wavelength scale at which the modal mapping evolves and measure the relative dispersion in the device, finding significant differences with idealised simulations. In addition to detailing the system, we also share the empirical mode transfer matrices, enabling future work in astrophotonic design, computational imaging, device fabrication feedback loops and beam shaping. 

\end{abstract*}


\section{Introduction}
Photonic lanterns~\cite{leon-saval_multimode_2005, leon-saval_photonic_2013, birks_photonic_2015} are mode converting devices that feature an adiabatic transition between a multimode waveguide and multiple single-mode waveguides. This feature enables numerous applications, such as space division multiplexing in telecommunications~\cite{velazquez-benitez_scaling_2018}, joint wavefront sensing with imaging in astrophotonics~\cite{norris_all-photonic_2020, norris_optimal_2022, lin_real-time_2023, jovanovic_2023_2023}, beam shaping~\cite{milne_coherent_2023, chandrasekharan_high-throughput_2025} and computational imaging~\cite{choudhury_computational_2020}. In all applications, knowledge of the optical mode transfer matrix -- the mapping from the modes at one end of the device to the modes at the other -- is critical. This mapping is linear (i.e. it satisfies superposition and scalar multiplication) when using the complex electric field, however past approaches typically only characterise devices in intensity. Inferring the mode transfer matrix, is then an under-constrained non-linear problem resulting in worse generalisation outside of the characterisation. Data-driven approaches have also been employed, requiring solution of this under-constrained problem by fitting task specific behaviour, such as predicting atmospheric (pupil plane) aberrations from the single-mode intensities in wavefront sensing~\cite{norris_all-photonic_2020, lin_real-time_2023}. A deeper, physical understanding of the modal mapping in any given photonic lantern would not only provide stronger priors to reduce data requirements for data-driven approaches, but enable faster design to fabrication loops and more realistic development of future applications such as post injection coherent beam combination~\cite{kim_coherent_2024}.

Simulations offer a tempting means of recovering the modal mapping by numerically propagating the modes supported at one end of the device, employing algorithms such as Eigenmode expansion~\cite{gallagher_eigenmode_2003} or the beam propagation method~\cite{van_roey_beam-propagation_1981}. However the precision with which device geometry can be specified in the manufacturing process is much looser than the level that causes variations in the simulated transfer matrix~\cite{davenport_photonic_2021, rypalla_large-quantity_2024}, along with manufacturing limitations and imperfections. Our results quantify this discrepancy and serve as a warning for the limitations of simulation-only design for these devices.

The majority of previous work aimed at characterising photonic lanterns has measured the intensities of modes, typically in monochromatic light~\cite{zhao_design_2024, yu_mode-dependent_2016, becerra-deana_fabrication_2024}. This is blind to the rich phase encodings and mode evolution with wavelength known to be present from theory and from success in previous wavefront sensing work. By measuring the coupling map (that is, the intensity of the single mode ports subject to tip/tilt for the beam incident on the multi-mode end) over many wavelengths~\cite{kim2024spectral} is able to infer the amplitude of the transfer matrix and some information about the modal dispersion of a 3 mode device. Whilst effective for few mode devices, the method would struggle for devices with more modes. Our approach instead directly measures the amplitude and phase of the electric field using digital off-axis holography, and is able to follow the evolution of the modes (including dispersion) in wavelength over a 73\,nm bandwidth. 

Holography~\cite{gabor1948new} refers to a wide range of techniques that typically measure the interference between the waves from an object and a reference wave, encoding both the amplitude and phase of the light from the object. Digital off-axis holography~\cite{zhang2021review} has been used in a range of applications, including spatial biphoton states~\cite{zia_interferometric_2023} and mode sorters/multiplexers~\cite{carpenter_all_2012,fontaine2019laguerre,fontaine_hermite-gaussian_2021}, and involves interfering the object and reference on a digital detector. Other work on photonic lanterns~\cite{van2022optical,xin_laboratory_2024} has also used digital off-axis holography, though only at a single wavelength, on a photonic lantern supporting a smaller number of modes. In addition to using a range of wavelengths and two polarisations, the characterisation system in this work deals with a \gls{MCF} fed photonic lantern, injecting light into an individual port with a \fib{} alignment stage. Through a \fib{} bundle as an adapter, the system can still characterise pigtailed devices. This work also publishes the results of the characterisation and draws generalisable insights about the behaviour of these devices.

There is also ongoing work to measure the transfer matrix of a photonic lantern~\cite{eikenberry_photonic_2024, romer_broadband_2025} proposing to use off-axis holography for the initial estimation of the modes and then using a precisely calibrated \gls{SLM} to form the required field at the multimode end of the device, probing each mode for characterisation. Our alternative approach removes the complexity of requiring precisely known input electric fields at the multimode end, instead demanding precise measurement of output electric fields: arguably a simpler problem that can be solved more accurately. Ultimately, characterisation in both directions is needed to test fundamental questions such as ``to what extent are lantern devices adiabatic?''.

\subsection*{Proposed approach}

The key contributions of this work include the design and implementation of a laboratory system for characterising a photonic lantern using digital off-axis holography. We share the testbed architecture, insights, lessons learned, code, and post-processed data. As part of this effort, we characterise a 19-port photonic lantern and compute its mode transfer matrix across different polarisations and wavelengths, offering initial insights into the modal mappings. Additionally, we identify trends in the coherence of the fields associated with different modes, including a direct measurement of differential modal dispersion. All code, and post-processed results are made available at
\href{https://github.com/ataras2/illuminating-the-lantern}{this page \faicon{github}}. 
For further technical details on the characterisation system, see Supplement 1.


\begin{figure}[ht]
    \centering
    \includegraphics[width = 0.85\textwidth]{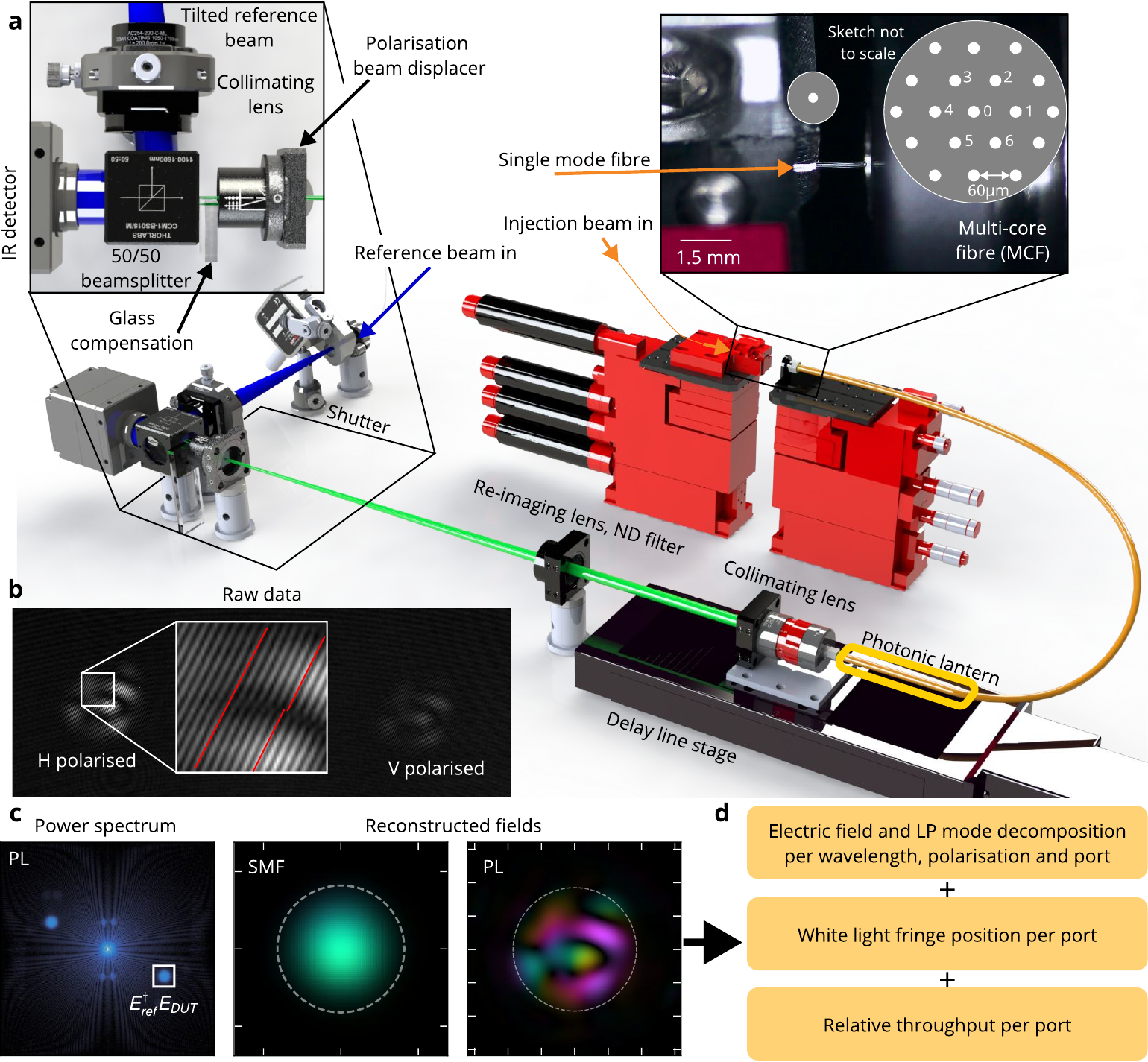}
    \caption[Visual overview of photonic lantern characterisation]{\textbf{Visual overview of photonic lantern characterisation}. \textbf{a} Labelled render of the laboratory setup. Light from a wavelength sweeping source is split into a reference (blue) and injection (orange) fibre. The injection beam excites a single port of the photonic lantern, producing a mode field that is re-imaged on the detector, with interference from a tilted reference beam producing the hologram. One arm of the interferometer has a delay line in order to maintain coherence during the wavelength sweep. \textbf{b} An example frame of raw data from the detector. The beam displacer splits the mode field into H and V polarisations. Fringes crests (red lines) reveal how the hologram captures phase information. \textbf{c} In post-processing, each polarisation is cropped (not shown), the coherent component in Fourier space (white box) extracted, and the electric field of the reference beam is removed. The result is the electric field of the device under test, either a single mode \fib{} (SMF) for validation or a photonic lantern (PL). \textbf{d} By taking these measurements with both a broadband source and a sweeping laser, the system characterises the port to electric fields and LP mode mapping, white light fringe position and relative throughput of the device under test.}
    \label{fig:overviewfig}
\end{figure}

A visual summary of our approach is shown in \autoref{fig:overviewfig}. Either a broadband or wavelength sweeping source is split into a reference and injection beam that each pass through a linear polariser before entering the optics shown. The injected beam is aligned to a single core of the \gls{MCF} (i.e. a single port of the photonic lantern), where the resulting mode is re-imaged onto the detector through a polarisation beam displacer, providing two orthogonal polarisations. The reference beam is collimated, tilted, rotated (such that the polarised axis is at 45$^\circ$ to the beam displacer axes) and projected onto the same detector, producing fringes required for digital off-axis holography. A broadband source is used to drive the delay line stage to the path length matched position for each port. Data is taken using the wavelength sweeping source over all ports, both with and without the reference beam. In post-processing (\autoref{fig:overviewfig} \textbf{c}), we recover the coherent component in Fourier space and remove the contribution of the reference beam, inferring the complete electric field of the device under test. Our pipeline then decomposes this into linearly polarised (LP) modes, providing the first ever multi-wavelength, polarisation decomposed characterisation of a photonic lantern in addition to revealing broadband coherence behaviour and relative throughput. 
{In this work, electric fields are visualised on a two dimensional colormap, with the phase axis selected to be perceptually uniform (the rate of color change with phase appears constant) and uniform in lightness (all phases appear equally bright) using the \textit{phase} colormap in \texttt{cmocean}~\cite{thyng2016true}, and the amplitude reflected in the value. Thus, pixels with the same amplitude appear equally bright irrespective of phase, which is not the case with the \textit{HSV} colormap that is typically used.}

\subsection*{Definition of the mode transfer matrix}
In this section, all coefficients and matrices are functions of wavelength and polarisation, however this is omitted for brevity.

A photonic lantern is a mode sorting device that maps from $n_{\mathrm{mm}}$ modes supported at the multimode end with complex coefficients $c_{\mathrm{mm}}\in \mathbb{C}^{n_{\mathrm{mm}}}$ (known as the `input' in e.g. wavefront sensing work, but the `output' in beam shaping work), to the $n_{\mathrm{sm}}$ modes supported at the single-mode \gls{MCF} or multiple \gls{SMF} end with coefficients $c_{\mathrm{sm}}\in \mathbb{C}^{n_{\mathrm{sm}}}$. These coefficients form the electric field in each respective basis, with
\begin{align}
    \bm{E}_{\mathrm{mm}} &= \sum_{i} c_{\mathrm{mm},i} \bm{E}^b_{\mathrm{mm},i},\ \ 
    \bm{E}_{\mathrm{sm}} = \sum_{i} c_{\mathrm{sm},i} \bm{E}^b_{\mathrm{sm},i},
\end{align}
where the sums are taken over all supported modes, $\bm{E}^b_{\mathrm{mm},i}$ is the $i$-th basis mode supported at the multimode end (assumed to be a LP mode in this work), and $\bm{E}^b_{\mathrm{sm},i}$ is the $i$-th basis mode at the \gls{MCF} end (a simple single-mode located at a position on a grid, typically hexagonal). Most photonic lanterns are designed such that $n_\mathrm{mm} =n_\mathrm{sm}$ at the shorter wavelengths of interest, but this doesn't need to be the case over the full bandwidth. Our approach makes no assumptions about either  $n_\mathrm{mm}$ or $n_\mathrm{sm}$, however we note that we expect oversampled devices ($n_\mathrm{mm} <n_\mathrm{sm}$) to lose light to cladding modes in the characterisation direction. It could be possible to recover the mapping in this case through exploiting the differential coherence between cladding and guided modes, and this is left as future work.

A transfer matrix $T\in \mathbb{C}^{n_\mathrm{sm} \times n_\mathrm{mm}}$, then, is the linear transformation applied by the photonic lantern, relating the above sets of coefficients as
\begin{align}
    c_\mathrm{sm} &= Tc_\mathrm{mm}.
\end{align}
Whilst these operations are linear in the modal bases, typical measurement systems capture the intensity $| \sum_{i} c_i \bm{E}_i|^2$, which is non-linear. 

Each row of the transfer matrix determines the response of the multicore \fib{} port to the electric field at the multimode end. In this work, we measure $T^\dagger$, the Hermitian of $T$. We inject a single mode at the multicore \fib{} end and use digital off-axis holography to measure the complex electric field at the multimode end, known as a principal mode~\cite{kim_coherent_2024}. Between measurements of different ports, the system drifts in phase and hence there is an ambiguous relative phase between different ports. Thus each reconstructed field contains a superposition with
\begin{align}
    c_\mathrm{mm} &= c_\mathrm{sm} T^\dagger = e^{i\phi_e} \bm{e}_j T^\dagger = e^{i\phi_e} T^\dagger_j,
\end{align}
where $\bm{e}_j$ is a vector of zeros with a one at index $j$, and $\phi_e$ is a (unknown) relative phase between ports. In other words, we measure each row of the transfer matrix up to a single absolute phase ambiguity. In the wavelength dimension, the quantities are measured coherently and rapidly so that there is no information-destroying phase wrapping nor large drift that occurs between samples. Hence the relative phase of a principal mode as a function of wavelength is known. 

\section{Methods}

\begin{figure}[h]
    \centering
    \includegraphics[width=0.9\linewidth]{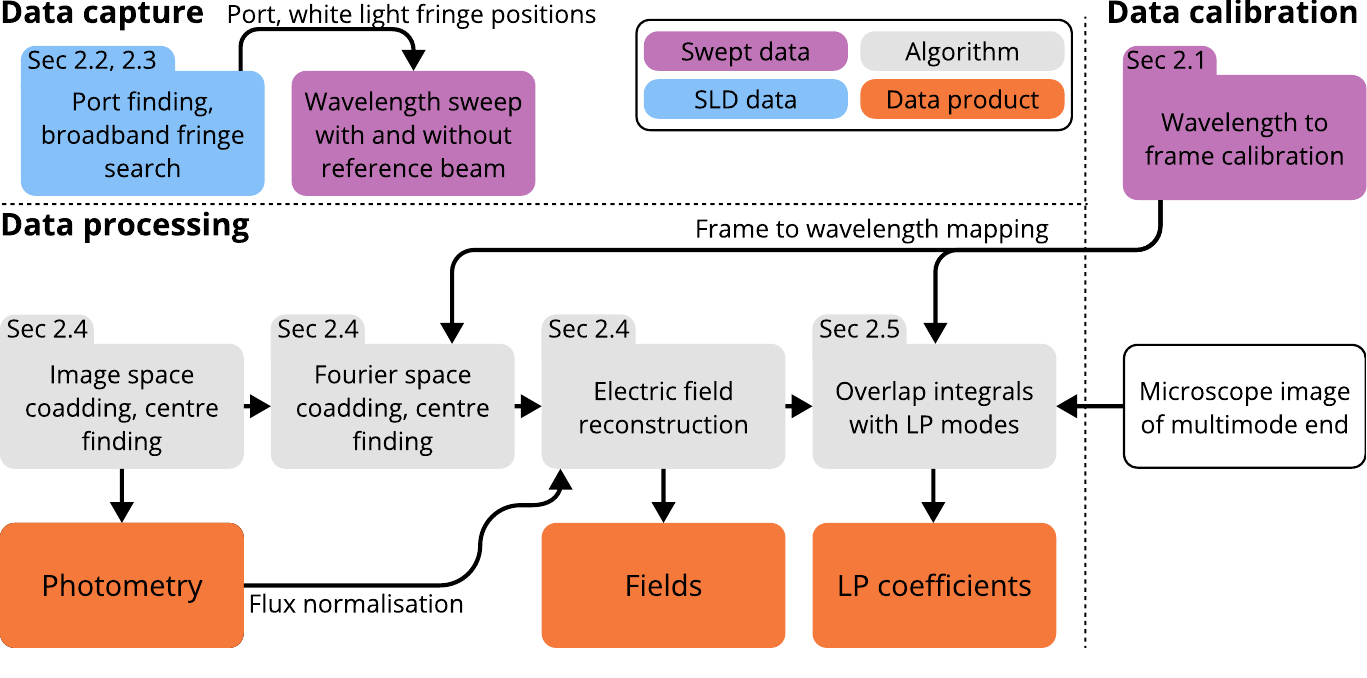}
    \caption{\textbf{Data capture and analysis block flowchart.} Numbered tabs show the relevant section in the main text. 
    }
    \label{fig:processing_block_diagram}
\end{figure}

\autoref{fig:processing_block_diagram} illustrates the interplay between data capture, post-processing and data products, with relevant steps linked to subsections. Data is first captured from illumination with a broadband \gls{SLD} source, with the port positions and the position of zero path delay found. These values are then used for the holographic data, which is captured while sweeping the source wavelength. In the processing pipeline, centres of the mode field in each polarisation are found in image space, and the off-axis component is fit in Fourier space. In each case, a smooth function is fit against wavelength. The data taken without the reference beam is used for photometry. The electric field is then reconstructed using the centres found previously. Finally, the projection onto LP modes is computed with overlap integrals, forming the transfer matrix that describes the modal mapping of the device. Further figures and details on the methods, including a system block diagram and fitting performance figures can be found in Supplement 1, Sec. 1. 


\subsection{Wavelength calibration}

The use of an electronic trigger system guarantees that the images captured during a wavelength sweep are consistently exposed through the same range of wavelengths, however the exact mapping must be calibrated. This frame index to wavelength calibration is done in two steps. First, the functional form of $\lambda(t)$ is fit using data from a wavemeter during the sweep. We find the sweep is weakly non-linear -- with a quadratic term present that causes the sweep speed to decrease from an initial value of 105\,nm/s to a final value of 90\,nm/s. Next, we mark the start and end of the sweep by monitoring the phase of the off-axis component. This is shown in Supplement 1, Figure S2. The total duration of the sweep is less than 1\,s. The calibrated mapping is then used for all later sweeps, and the laser calibration routine is run at the start of every set of measurements to ensure no drifts. In short, 78 samples are captured spanning the range of 1507.5\,nm to 1580.5\,nm.

\subsection{Exciting a single port}

We excite a single port (i.e. one principal mode) of the photonic lantern from the multicore end via butt coupling on a motorized \fib{} alignment stage. 
Using multicore fibres creates the additional challenge that we need to align fibres rather than simply using a fibre switcher. We note that any mismatch between the mode field of the input \fib{} and the \gls{MCF} causes a loss in power, however this does not affect the characterisation of the transfer matrix as we do not need to measure the absolute throughput (which is better measured by other methods). We know the falloff with respect to position is Gaussian~\cite{marcuse_loss_1977}, so we measure the total intensity in the images at a given motor position and monitor how this changes. The flux is then fit as the sum of Gaussian functions, with centres lying on a hexagonal grid. We fit an origin, separation and rotation of the grid, as well as the nuisance parameters of the mode field diameter and the individual throughput per port. For a visualisation, see Supplement 1, Figure S3.

\subsection{White light fringe finding}
Path length differences between the two arms result in a phase measurement that changes as a function of wavelength. In order to capture the evolution of the mode transfer matrix with wavelength coherently, the phase of the measured field must change by less than $\pm\pi$ between frames. The \textit{white light fringe} refers to the case where this phase change is virtually zero over a wide range of wavelengths, such that interference fringes are visible even at large bandwidth. The position of the white light fringe in the interferometer is found by changing the length of the injection arm using the delay line stage shown in \autoref{fig:overviewfig} \textbf{a} and a broadband (50\,nm) \gls{SLD}. The white light fringe occurs when the off-axis power is maximised. The position used when taking data with the sweeping source is the midpoint between the maxima for the H and V polarisations, such that both are high visibility during the sweep. Results from this step are shown in Supplement 1, Figure S6. 


\subsection{Digital off-axis holography reconstruction}



In order to reconstruct the field, we must first centre in the image plane. For each polarisation and at each wavelength, the photometry data is co-added, the outline of the core found through Canny edge detection~\cite{canny2009computational} and then the centre estimated using a least squares circle estimator~\cite{jekel2016obtaining}. A sub-pixel Fourier shift of the frame then places the pattern at the centre and crops the image. In the Fourier plane, we use a custom geometric circle finding algorithm that seeks the smallest disk to encase a given amount of flux. Both the image and Fourier centre estimates are smoothed by fitting as a function of wavelength, reflecting weak chromaticity in system components (in image space) or the radial frequency smearing (in Fourier space). For a visualisation of these steps, see Supplement 1, Figure S4.

\newcommand{\inner}[2]{\left\langle #1, #2 \right\rangle}

The reconstruction of the electric field $\bm{E}_{\mathrm i}$ from an image at a known wavelength is given by
\begin{align}
    \bm{E}_{\mathrm i} &= \bm{E}_{\mathrm{ref}}'\mathcal{F}^{-1} (\bm{T}_\theta(\bm{W}_{\mathrm{Fourier}}) (\mathcal{F} (\bm{W}_{\mathrm{image}}\bm{I}_{\mathrm i}))),
\end{align}
where $\bm{E}_{\mathrm{ref}}$ is the estimated electric field of the reference beam, $'$ denotes the conjugate, $\mathcal{F}$ denotes the Fourier transform (with inverse $\mathcal{F}^{-1}$), $\bm{T}_\theta$ is a translation operator with parameter $\theta = (\theta_u, \theta_v)$ (implemented using a phase ramp in Fourier space), $\bm{W}_{\mathrm{image}}$ and $\bm{W}_{\mathrm{Fourier}}$ are windows with width and shape parameters, and $\bm{I}_{i}$ is the holographic image from port $i$. The structure of the Fourier space is shown in \autoref{fig:overviewfig} \textbf{c}. The electric field of the reference $\bm{E}_{\mathrm{ref}}$ is estimated with the phase determined by a Fourier transform of a delta function located at the peak of the power spectrum, and a uniform amplitude.

The filtering of only off-axis components means that low order aberrations manifesting near zero spatial frequency are removed, and results in strong robustness to read noise. Intuitively the reconstruction of off-axis holography data can be thought of as using the intensity to recover the field amplitude and the curvature of the fringes to infer the derivative of the phase.

\subsection{Projection onto LP modes}

After reconstructing the fields, we calculate the overlap integral $\eta$, given by
\begin{align}\label{eqn:holo_reconstruct}
    \eta_{i,j} &= \inner{\bm{E}_{\mathrm i}}{\bm{E}^{b}_{\mathrm{mm},j}},\\
    \text{where  } \inner{\bm{A}}{\bm{B}} &=         \frac{\int \bm{A}\bm{B}^* \mathrm{d}A}{\sqrt{\left(\int \left| \bm{A} \right|^2 \mathrm{d}A \right)\left(\int \left| \bm{B} \right|^2 \mathrm{d}A\right)}},
\end{align}
where the integrals are over the plane perpendicular to the fibre face, $i$ is the index of the lantern port used and again $\bm{E}^b_{\mathrm{mm},j}$ is the $j$-th basis mode supported at the multimode end. A normalised field is denoted as $\hat{\bm{E}}$. We approximate the integrals as a sum over a grid, with mode profiles found using the \texttt{ofiber} package~\cite{ofiber}.
The physical fibre diameter used to generate the LP modes is measured using a microscope beforehand, and the refractive index contrast is known from manufacturing specifications. The plate scale of the injection arm is fit such that the maximum (power weighted) overlap value is achieved. 

In this work we only employ LP modes, however we note that this is only an approximation. Due to structures from the multicore \fib{} remaining in the multimode end, the refractive index is not a simple top hat function. Future work could directly apply basis modes that are solved numerically to better reflect this structure. 

\section{Results}

Here we present the results of the characterisation of the 19-port photonic lantern. See Supplement 1, Sec. 3 for validation of the characterisation system on a \gls{SMF}, white light fringe finding and throughput/overlap integral performance of the device. 

We fabricate a 19-port photonic lantern with a multimode diameter of 34.7\,$\mu$m, a core refractive index of 1.44 and a cladding index with a contrast of $5.5\times10^{-3}$. Over the wavelengths swept, the output supports 23 LP modes below $\lambda_c = 1562$\,nm and 21 LP modes above $\lambda_c$. Hence, this is a marginally undersampled lantern, manufactured in order to guarantee that all the modes of interest are strongly bound over the whole operational wavelength range.

The characterisation system takes data for this device in just over one hour, with time primarily spent on finding the position of the white light fringe. The data capture with the swept wavelength source takes less than 10 minutes, an important feature in enabling a tight feedback loop in the manufacture of devices. 

\subsection*{Reconstruction visualisations}

\begin{figure}[h]
    \centering
    \includegraphics[width=0.99\linewidth]{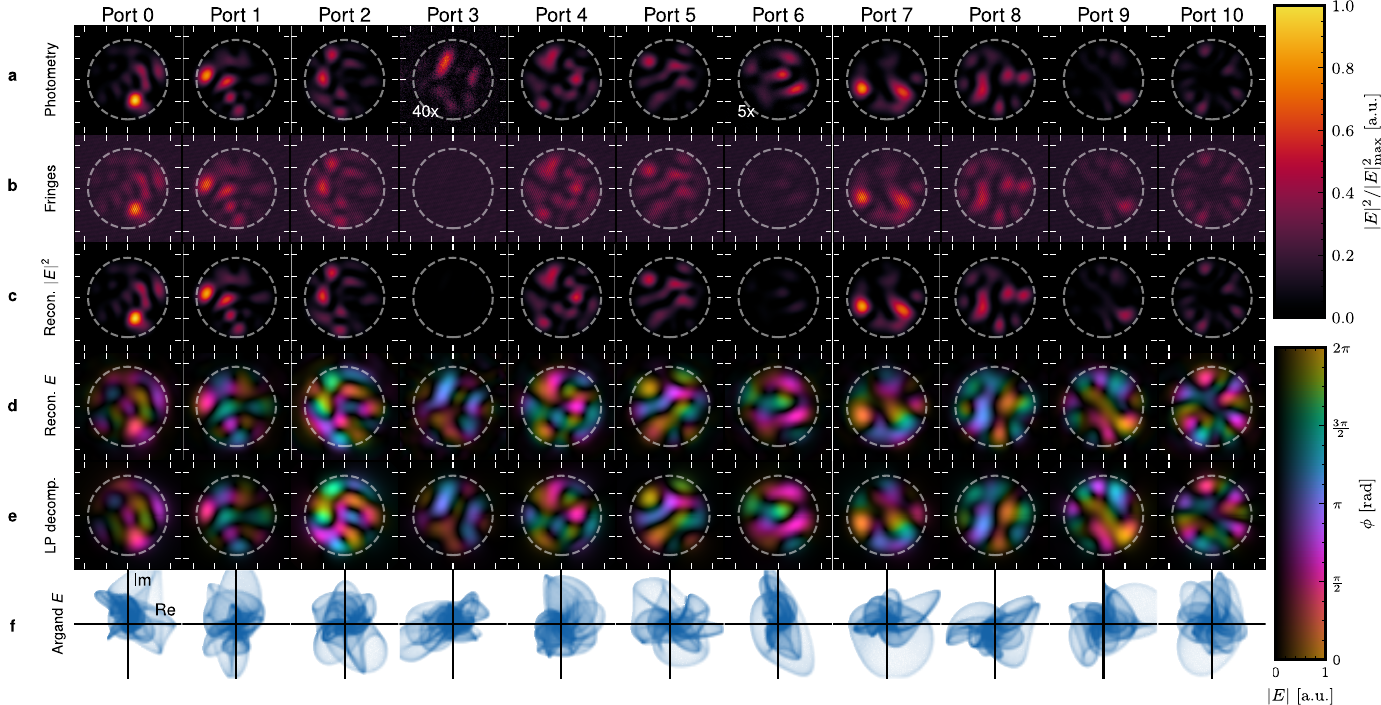}
    \caption{\textbf{Data and reconstruction at a single wavelength and polarisation.} We capture two sets of data: \textbf{a} photometry and \textbf{b} fringes. Each row is normalised. We only show data at $\lambda=1513.9$\,nm and in H polarisation, so ports with power in V (e.g. port 3, stretched by a factor 40 in photometry) appear dim. The distance between the white ticks is 10\,$\mu$m and the circle indicates the core size of the multimode end. \textbf{c} The reconstruction has an intensity that is similar to the photometry. \textbf{d} The reconstructed electric field is well recovered, even at low signal such as port 3. \textbf{e} The decomposition into LP modes matches the reconstruction well. \textbf{f} Visualisation of the fields in an Argand diagram, where each pixel from \textbf{d} is plotted as a point. While some electric fields appear qualitatively similar in \textbf{d}, \textbf{f} highlights the diversity of the modes supported. See Visualization 1 for a video of all ports through all wavelengths.}
    \label{fig:fringes_fields_and_phasors_cbar}
\end{figure}

\autoref{fig:fringes_fields_and_phasors_cbar} shows the steps of our reconstruction. The first two rows are dark subtracted data, taken without and with the reference beam respectively. Despite each port having the same input beam (including polarisation), after propagation through the device the polarisation changes in a port dependent way. In the cases where a port appears dim in H (such as port 3), there is good signal in V. The following rows show our reconstruction. The intensity of the reconstructed field $|E|^2$ is in good agreement with the photometry (reference beam blocked) data. The reconstructed fields show localised blobs with nulls between them, a wide range of phases and are well contained by the core outline, as expected. The fields are also consistent with a superposition of LP modes, with the decomposition showing agreement to the field itself. This decomposition is reasonable but imperfect (see e.g. the bottom right of port 4), motivating future work with a more complex modal basis. Finally, we also visualise the fields on an Argand diagram, illustrating a unique structure for each port. Importantly, the reconstruction is of reasonable quality even in low SNR -- with the resulting field still well explained by LP modes. This is due to the robustness of digital off-axis holography where only the coherent information is extracted, and hence contamination such as background and read noise are suppressed.

\subsection*{Slicing the transfer matrix}

\begin{figure}[h]
    \centering
    \includegraphics[width=0.99\linewidth]{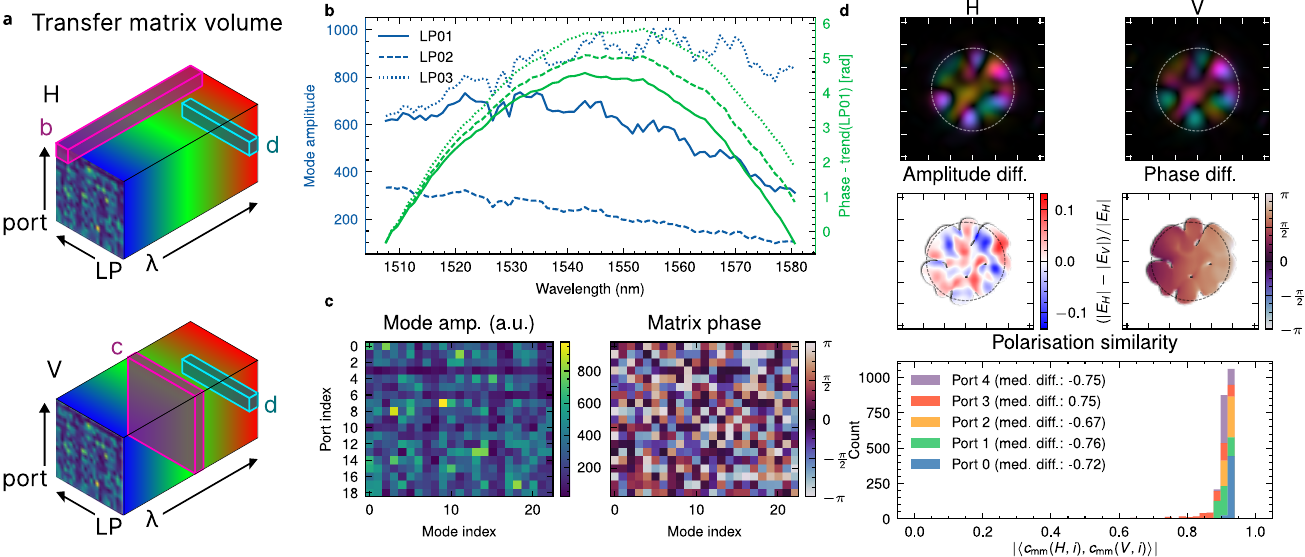}
    \caption{\textbf{Slices through the transfer matrices of the 19-port lantern}.
    \textbf{a} Our characterisation system produces transfer matrix cubes that are coherent through wavelength. Here we explore three slices through the cube. \textbf{b} Mode evolution along wavelength. \textbf{c}~Monochromatic transfer matrix per polarisation. This lantern has contributions from many modes to each port, which enables good inference of the input field with only intensity measurements. \textbf{d} 
    Slices of the same port and wavelength in different polarisations. The fields appear visually similar (top row), however subtle differences exist (middle row). Over ports and wavelengths the fields of different polarisations (stacked histogram, bottom row) exhibit a dissimilarity that is not explained by a difference in signal alone (see text for details).}
    \label{fig:transfer_matrix_full}
\end{figure}

\autoref{fig:transfer_matrix_full} \textbf{a} summarises the measurement dimensions available in our post processed transfer matrices. Slices through wavelength (\autoref{fig:transfer_matrix_full} \textbf{b}) show the evolution of modes in amplitude and phase. There are some oscillations present due to birefringence in the system (see Supplement 1 for details), but they are common mode and the trend of modes is clearly a slowly varying function in both amplitude and phase. Relative measurements are very precise and explored further in \autoref{fig:modal_dispersion}. Slices at a particular wavelength (\autoref{fig:transfer_matrix_full} \textbf{c}) yield monochromatic transfer matrices in a given polarisation. Previous work~\cite{norris_optimal_2022} has only ever used simulations to generate these, and hence shows strong symmetry between ports on opposite sides of the multicore \fib{} grid. We do not observe this, highlighting the significance of manufacturing variations from a target design.

Furthermore, we can compare slices in polarisation, visualised in \autoref{fig:transfer_matrix_full}\,\textbf{d}. The electric fields from each polarisation appear identical up to a spatially constant phase offset. Taking differences in amplitude, however, we see changes by up to 10\% in some regions, with a structure that is not consistent with centring residuals alone. In phase, we observe a near uniform difference, with a weak tilt likely caused by a minor deviation in angle from the beam displacer. In the bottom row, we report a similarity statistic over a few ports and all wavelengths in a stacked histogram. We normalise the LP coefficients for a given port and wavelength, and then compute the magnitude of the inner product between the two polarisations. For very similar fields with a arbitrary, spatially constant phase offset, this metric is close to unity. Since the quality of the LP overlap is related to the flux, we also report the median flux difference between H and V, i.e. the median of $(f_H-f_V)/(f_H+f_V)$ over all wavelengths. Values closer to $\pm1$ have less flux in one polarisation, are expected to have less correct LP mode projections, and hence lower similarity even if the true fields are similar. We observe a port independent bound of around 0.94, consistent with the phase and amplitude differences shown above that are from the limitations of the imaging system, such as the tolerance on the parallelism of the polarisation beam displacer and the compensating plate. There is, however, port dependence that indicates a weak difference in the electric fields of different polarisations. This is most evident between ports 0 and 2, from which a LP quality argument would favour higher similarity in port 2 but the data suggests port 0 has higher similarity. This warrants further investigation with more devices.

\subsection*{Wavelength dependence}

\begin{figure}[h]
    \centering
    \includegraphics[width=0.85\linewidth]{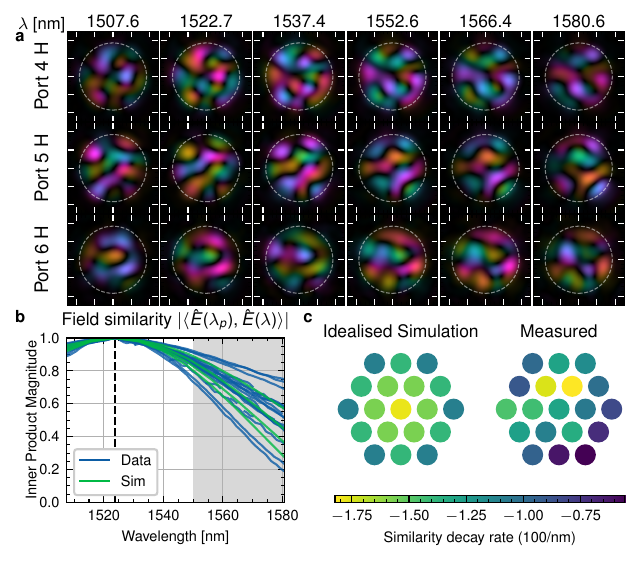}
    \caption{\textbf{Wavelength evolution of modes in a photonic lantern.} 
    \textbf{a} Electric field plots over ports 4-6 over the full bandwidth. 
    \textbf{b} Field similarity between a fixed wavelength $\lambda_p$ (black dashed line) and all other wavelengths within the same port, for data and simulation. Each curve is one port. The falloff becomes near linear after $20\,$nm (grey box). \textbf{c} Gradient of similarity decay for simulated and measured fields over all ports.}
    \label{fig:field_with_wavel}
\end{figure}

\autoref{fig:field_with_wavel} shows the breadth of wavelength dependent data available from our system. First, we show the electric field for one polarisation for different ports. All ports change at approximately the same rate, with neighbouring panels appearing identical up to an overall phase, while distant panels are clearly different. The overall phase evolution is generated from being offset from the white light fringe, recalling that we took the midpoint between the two polarisations and hence are $\sim85\,\mu$m from the white light fringe in both. 
Next, we quantify the wavelength scale of field evolution in the device by measuring the overlap integral of the field at a particular wavelength (black dashed line) compared to all other wavelengths. The characteristic scale for the mode field to change noticeably is of order 20\,nm, broadly consistent with previous results in wavefront sensing~\cite{norris_all-photonic_2020}. The falloff is near linear around 20\,nm from the fixed wavelength. Both the simulation and measurement agree in the functional form (smooth turning point with linear decay) but the rate of decay can be slower or faster in the actual device, depending on the port. Finally, we visualise the rate of decay in this linear region on the port map of the device. Generally, ports towards the centre of the \gls{MCF} decay faster, however the device clearly does not have the symmetry of the idealised simulation. The difference in wavelength needed for principal modes to become orthogonal could be as little as $100/1.8 = 56\,$nm (3.6\% bandwidth) or as much as $100/0.6 = 167\,$nm (10.7\% bandwidth).

\subsection*{Direct measurement of relative modal dispersion}
Finally, since our characterisation system probes the device response to coherent light, we are able to directly measure differential observables of the modal dispersion of the photonic lanterns. From data such as \autoref{fig:transfer_matrix_full}~\textbf{b}, we observe that oscillations in measured phase are common between modes, suggesting that these are present from weak birefringence outside of the photonic lantern itself. Hence, our approach is much more sensitive to differential measurements between modes than at first glance. We now consider differential phase measurements between modes to directly measure the relative modal dispersion, noting that this result is insensitive to any birefringence in the system and the \gls{MCF} part of the photonic lantern. The modal dispersion is significant for many applications, dictating the bandwidth over which systems using the photonic lantern can maintain coherence.

\begin{figure}[h]
    \centering
    \includegraphics[width=0.8\linewidth]{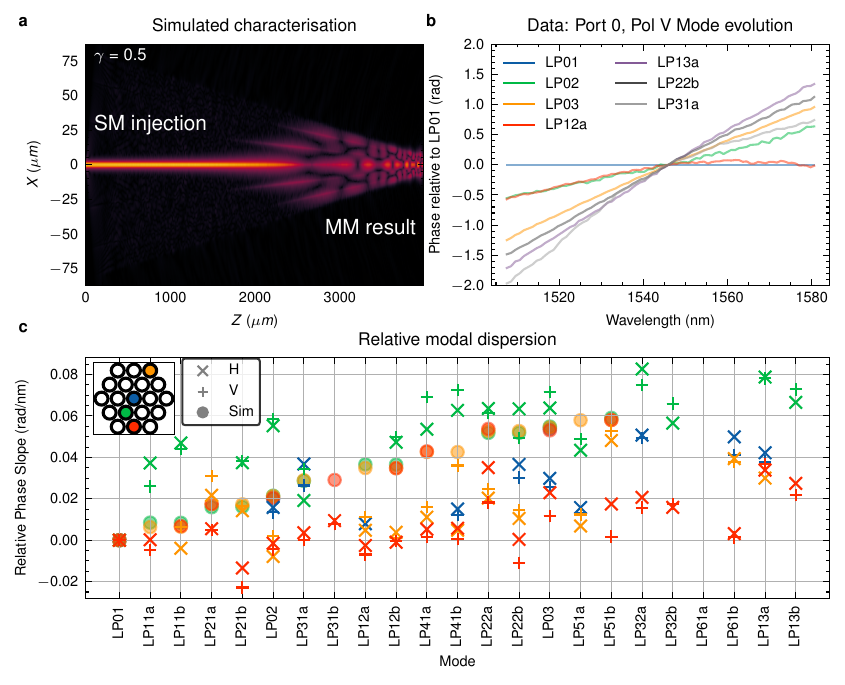}
    \caption{\textbf{Relative modal dispersion measurements of the photonic lantern.} 
    \textbf{a} Cut through of a photonic lantern simulation, showing amplitude of the electric field when exciting the central port. \textbf{b} Differential phase of a few modes relative to LP01 in a single port, with offsets for visualisation. Evolution is nearly linear, as expected.  
    \textbf{c} Gradient of the phase slope in \textbf{b} for every mode, in both polarisations for the ports indicated. A cut is taken to only include modes with all amplitudes above 80 data units to avoid low signal contamination. Errorbars (virtually invisible) show standard deviation over six sweeps. } 
    \label{fig:modal_dispersion}
\end{figure}

\autoref{fig:modal_dispersion} shows this analysis with a comparison to an idealised, yet typically used, scalar simulation. For each mode, we fit a line to the phase relative to LP01, the gradient of which is the relative modal dispersion. This is repeated for different polarisations and with the simulated data and summarised in \autoref{fig:modal_dispersion} \textbf{c}. Each measurement is extremely repeatable, with the median scatter (standard deviation from 6 sweeps) over all modes above the cut as $10^{-4}$\,rad/nm. We can also estimate an upper bound for the systematic error introduced by background contamination at low signal. A 5\% background signal can make at most a $2\times\arctan(0.05)$\,rad contribution to the phase over the whole bandwidth, or a change $0.1/73=1.5\times10^{-3}$\,rad/nm in the signal plotted, an order of magnitude smaller than the differences observed between modes. We also note that this upper bound on systematic error would not apply to differences between ports, as this contamination is common to all ports. Hence, we conclude that majority of the scatter between ports and modes is real signal, and that the simulation is too idealised. There are also subtle differences between the polarisations that again warrant further investigation with different devices.

\section{Discussion}

In this section we highlight some implications of our findings on applications involving photonic lanterns. 

First, we provide a warning for the use of simulation alone in modelling these devices. Previous simulation results e.g.~\cite{norris_optimal_2022} have shown high symmetry in the transfer matrix between ports on opposite sides of the hexagonal grid. Our characterisation shows no evidence of this, highlighting the need for physical device characterisation over simulation alone. We would be capable of discerning if this variation is a fixed or variable offset from the design target given multiple devices, and leave this as future work. Broadly, simulations seem to match the same order of magnitude and functional form as the fabricated device -- such as with the gradient of similarity decay in \autoref{fig:field_with_wavel} or the modal dispersion in \autoref{fig:modal_dispersion} -- but the predictions of observables vary from measured performance in a significant way that could be ruinous if relied upon for downstream tasks. 

As a more concrete example, instrument concepts that involve coherent combination of photonic lantern outputs, such as~\cite{kim_coherent_2024} require careful consideration of the port dependent optical path difference induced by the multicore fibre (measured at up to 750\,$\mu$m per metre of \fib{}) as well as the modal dispersion. Our modal dispersion findings (\autoref{fig:modal_dispersion}) suggest that such an instrument must be spectrally dispersed with a sampling finer than 1.6\,nm (for 0.1\, rad difference) to maintain coherence between different modes in different ports. We also find that the typical wavelength scale for the evolution of the principal mode is of order 20\,nm, but this value varies between ports.

Next, data driven approaches, especially where data is expensive, should leverage characterisation results as a prior to fitting. For example, an image reconstruction algorithm for high angular resolution astrophotonics should use the measured transfer matrix as the heart of a model that can be fine tuned in-situ using intensity measurements. More generally, estimators using the characterised photonic lantern should be regularised to be consistent with the characterisation. The transfer matrix can also serve in optimal experiment design, where observations could be planned around maximising sensitivity to scientific outcomes depending on the location of objects in the device's field of view.

Previous work for computational imaging~\cite{choudhury_computational_2020} with photonic lanterns for areas such as microendoscopy has promising initial results. In particular, more advanced and powerful imaging schemes leverage coherent combination, and hence require knowledge of the relative phases and amplitudes the fields induced by exciting each port. Our characterisation system provides almost all of this information directly. For microendoscopy specifically, the mode/port counts must be considerably higher than the 19-port device in this work, however the approach can scale whilst still remaining tractable -- we envisage the time needed for swept data taking to be around 10s per port, equivalent to characterising a 1000 port device in under 3 hours. In this case, time taken to find the white light fringe could be decreased significantly by using a bandpass filter for a longer coherence length (hence less sampling of the delay line motor) and a more innovative sampler that leverages correlation between ports (rather than a fixed step size). 

Beyond immediate applications, the existence and extent of birefringence and polarisation dependent loss within photonic lanterns remain open questions; weak differences between fields with different polarisations are present, and dispersions are different, warranting further investigation with a wider range of devices. Our characterisation approach offers sensitivity to these quantities, however upgrades to reduce system birefringence might be required to make convincing progress. Another possible upgrade would be to vary the input polarisation at device injection in addition to the existing measurement of output polarisation. Such an experiment would verify to what extent orthogonal polarisations remain orthogonal through the device, with consequences for polarisation multiplexing in telecommunications.

Our system has the limitation that we can not directly measure the relative phase between principal modes of different ports since the laboratory environment causes the two arms to drift in phase between measurements of different ports. However, such a measurement is very sensitive to the bend of the multicore fibre that feeds the device (see Supplement 1, Figure S6). If the application after characterisation requires coherent combination -- where the phase between ports matters -- then the system is sensitive to the environment near the multicore fibre and should have this aspect characterisable in-situ. 

\section{Conclusion}
Photonic lanterns provide an efficient means of mapping from a multimode basis set to multiple single modes (or vice versa), however this mapping is only loosely constrained at the design stage, with manufacturing limitations causing significant deviations from the target. In this work, we design and implement a characterisation system that directly measures the complex mode transfer matrix, using digital off-axis holography to reconstruct the electric field of the resulting mode from injection into a single port. We follow the evolution of these fields over a 73\,nm range near 1550\,nm. This provides the first multi-wavelength, polarisation decomposed characterisation of these multimode converters. In analysing the measured transfer function, we uncover the typical wavelength scale of principal mode evolution (agreeing with inferences made in previous work) and make a direct measurement of the modal dispersion within the device. We find differences in dispersion and symmetries compared to idealised simulation, reinforcing the need for empirical characterisation of all manufactured devices. The system presented and the resulting data enhance future work in astrophotonic design, computational imaging, device fabrication and beam shaping.

There are exciting avenues of future work with this characterisations system. It is now possible to study how the transfer matrix varies with different architectures, such as hybrid mode-selective photonic lanterns. Furthermore, the manufacturing repeatability of these devices can be measured directly. Finally, further work is needed to understand polarisation dependence in the photonic lantern, and how this changes with device design and fabrication parameters. 

\begin{backmatter}
\bmsection{Funding}
B. Norris is the recipient of an Australian Research Council Discovery Early Career Award (Grant No. DE210100953) funded by the Australian Government. S. Leon-Saval and C. Betters acknowledge support by the Air Force Office of Scientific Research under award number FA2386-23-1-4108.

\bmsection{Acknowledgment}
We would like to acknowledge the helpful discussions and insight from all our collaborators on the photonic lantern sensing effort, including Michael Fitzgerald, Olivier Guyon, Nemanja Jovanovic, Yoo Jung Kim, Manon Lallement, Jonathan Lin, Julien Lozi, Sebastien Vievard, Yinzi Xin, and others from California Institute of Technology, Univ. of California Los Angeles, Univ. of California Irvine, and the SCExAO team at the Subaru Telescope, NAOJ. A. Taras would also like to thank David Sweeney, Daniel Dahl, Benjamin Pope and Max Charles for insightful discussions about figures. 

Microsoft Copilot was used in developing the software for this work. We acknowledge support from Astralis -- Australia’s optical astronomy instrumentation consortium -- through the Australian Government's National Collaborative Research Infrastructure Strategy (NCRIS) program. 

We would like to acknowledge the Gadigal People of the Eora nation, the traditional owners of the land on which most of this work was completed.

\bmsection{Disclosures}
The authors declare no conflicts of interest.

\bmsection{Data Availability} All code, the measured optical mode transfer matrix and additional post-processed data can be found through 
\href{https://github.com/ataras2/illuminating-the-lantern}
{this GitHub page \faicon{github}} or with DOI: \href{https://doi.org/10.5281/zenodo.17815747}{10.5281/zenodo.17815747}. 

\bmsection{Supplemental document}
See Supplement 1 for supporting content. 
See \href{https://opticapublishing.figshare.com/articles/media/fringes_fields_and_phasors_mp4/30529982}{Supplement 2} for a video showing \autoref{fig:fringes_fields_and_phasors_cbar} over all cores and as a function of wavelength through the video time. 

\end{backmatter}
\bibliography{sample}

@misc{ofiber,
    author = "Prahl, S.",
    title = "ofiber: a python module for light propagation in optical fibers",
    year = {2024}
}

@article{zhang2021review,
  title={A review of common-path off-axis digital holography: towards high stable optical instrument manufacturing},
  author={Zhang, Jiwei and Dai, Siqing and Ma, Chaojie and Xi, Teli and Di, Jianglei and Zhao, Jianlin},
  journal={Light: advanced manufacturing},
  volume={2},
  number={3},
  pages={333--349},
  year={2021},
  publisher={Light: Advanced Manufacturing}
}

@article{kim2024spectral,
  title={Spectral characterization of a three-port photonic lantern for application to spectroastrometry},
  author={Kim, Yoo Jung and Fitzgerald, Michael P and Lin, Jonathan and Lozi, Julien and Vievard, S{\'e}bastien and Xin, Yinzi and Levinstein, Daniel and Jovanovic, Nemanja and Leon-Saval, Sergio and Betters, Christopher and others},
  journal={Journal of Astronomical Telescopes, Instruments, and Systems},
  volume={10},
  number={4},
  pages={045004--045004},
  year={2024},
  publisher={Society of Photo-Optical Instrumentation Engineers}
}

@phdthesis{jekel2016obtaining,
  title={Obtaining non-linear orthotropic material models for PVC-coated polyester via inverse bubble inflation},
  author={Jekel, Charles F},
  year={2016},
  school={Stellenbosch: Stellenbosch University}
}

@inproceedings{van2022optical,
  title={Optical field characterization using off-axis digital holography},
  author={van der Heide, Sjoerd and van Esch, Bram and van den Hout, Menno and Bradley, Thomas and Velazquez-Benitez, Amado M and Fontaine, Nicolas K and Ryf, Roland and Chen, Haoshuo and Mazur, Mikael and Antonio-L{\'o}pez, Jose Enrique and others},
  booktitle={Optical Fiber Communication Conference},
  pages={M3Z--6},
  year={2022},
  organization={Optica Publishing Group}
}

@article{fontaine2019laguerre,
  title={Laguerre-Gaussian mode sorter},
  author={Fontaine, Nicolas K and Ryf, Roland and Chen, Haoshuo and Neilson, David T and Kim, Kwangwoong and Carpenter, Joel},
  journal={Nature communications},
  volume={10},
  number={1},
  pages={1865},
  year={2019},
  publisher={Nature Publishing Group UK London}
}

@article{leon-saval_multimode_2005,
	title = {Multimode fiber devices with single-mode performance},
	volume = {30},
	copyright = {https://doi.org/10.1364/OA\_License\_v1\#VOR},
	issn = {0146-9592, 1539-4794},
	url = {https://opg.optica.org/abstract.cfm?URI=ol-30-19-2545},
	doi = {10.1364/OL.30.002545},
	language = {en},
	number = {19},
	journal = {Optics Letters},
	author = {Leon-Saval, S. G. and Birks, T. A. and Bland-Hawthorn, J. and Englund, M.},
	month = oct,
	year = {2005},
	pages = {2545},
}

@article{leon-saval_photonic_2013,
	title = {Photonic lanterns},
	volume = {2},
	issn = {2192-8614, 2192-8606},
	url = {https://www.degruyter.com/document/doi/10.1515/nanoph-2013-0035/html},
	doi = {10.1515/nanoph-2013-0035},
	language = {en},
	number = {5-6},
	journal = {Nanophotonics},
	author = {Leon-Saval, Sergio G. and Argyros, Alexander and Bland-Hawthorn, Joss},
	month = dec,
	year = {2013},
	pages = {429--440},
}

@article{van_roey_beam-propagation_1981,
	title = {Beam-propagation method: analysis and assessment},
	volume = {71},
	copyright = {https://doi.org/10.1364/OA\_License\_v1\#VOR},
	issn = {0030-3941},
	shorttitle = {Beam-propagation method},
	url = {https://opg.optica.org/abstract.cfm?URI=josa-71-7-803},
	doi = {10.1364/josa.71.000803},
	language = {en},
	number = {7},
	journal = {Journal of the Optical Society of America},
	author = {Van Roey, J. and Van Der Donk, J. and Lagasse, P. E.},
	month = jul,
	year = {1981},
	note = {Publisher: Optica Publishing Group},
	pages = {803},
}

@article{kim_coherent_2024,
	title = {Coherent {Imaging} with {Photonic} {Lanterns}},
	volume = {964},
	issn = {0004-637X, 1538-4357},
	url = {https://iopscience.iop.org/article/10.3847/1538-4357/ad245e},
	doi = {10.3847/1538-4357/ad245e},
	language = {en},
	number = {2},
	urldate = {2024-10-31},
	journal = {The Astrophysical Journal},
	author = {Kim, Yoo Jung and Fitzgerald, Michael P. and Lin, Jonathan and Sallum, Steph and Xin, Yinzi and Jovanovic, Nemanja and Leon-Saval, Sergio},
	month = apr,
	year = {2024},
	pages = {113},
}

@article{davenport_photonic_2021,
	title = {Photonic lanterns: a practical guide to filament tapering},
	volume = {11},
	copyright = {https://doi.org/10.1364/OA\_License\_v1\#VOR-OA},
	issn = {2159-3930},
	shorttitle = {Photonic lanterns},
	url = {https://opg.optica.org/abstract.cfm?URI=ome-11-8-2639},
	doi = {10.1364/ome.427903},
	language = {en},
	number = {8},
	urldate = {2025-07-23},
	journal = {Optical Materials Express},
	author = {Davenport, John J. and Diab, Momen and Deka, Pranab J. and Tripathi, Aashana and Madhav, Kalaga and Roth, Martin M.},
	month = aug,
	year = {2021},
	note = {Publisher: Optica Publishing Group},
	pages = {2639},
}

@inproceedings{rypalla_large-quantity_2024,
	address = {Yokohama, Japan},
	title = {On the large-quantity fabrication and reproducibility of all-fiber photonic lanterns},
	url = {https://www.spiedigitallibrary.org/conference-proceedings-of-spie/13100/3018846/On-the-large-quantity-fabrication-and-reproducibility-of-all-fiber/10.1117/12.3018846.full},
	doi = {10.1117/12.3018846},
	language = {en},
	urldate = {2025-07-24},
	booktitle = {Advances in {Optical} and {Mechanical} {Technologies} for {Telescopes} and {Instrumentation} {VI}},
	publisher = {SPIE},
	author = {Rypalla, Julian and Vješnica, Stella and Madhav, Kalaga V. and Lorenz, Adrian and Eschrich, Tina and Schmälzlin, Elmar and Dinkelaker, Aline N. and Roth, Martin M.},
	editor = {Navarro, Ramón and Jedamzik, Ralf},
	month = aug,
	year = {2024},
	pages = {254},
}

@article{zhao_design_2024,
	title = {Design and characterization of a self-matching photonic lantern for all few-mode fiber laser systems},
	volume = {32},
	issn = {1094-4087},
	url = {https://opg.optica.org/abstract.cfm?URI=oe-32-10-16799},
	doi = {10.1364/OE.520588},
	language = {en},
	number = {10},
	urldate = {2025-02-28},
	journal = {Optics Express},
	author = {Zhao, Li and Li, Wei and Chen, Yunhao and Yu, Ting and Zhao, Enming and Tang, Jianing},
	month = may,
	year = {2024},
	pages = {16799},
}

@article{yu_mode-dependent_2016,
	title = {Mode-dependent characterization of photonic lanterns},
	volume = {41},
	copyright = {https://doi.org/10.1364/OA\_License\_v1\#VOR},
	issn = {0146-9592, 1539-4794},
	url = {https://opg.optica.org/abstract.cfm?URI=ol-41-10-2302},
	doi = {10.1364/OL.41.002302},
	language = {en},
	number = {10},
	urldate = {2025-03-03},
	journal = {Optics Letters},
	author = {Yu, Dawei and Fu, Songnian and Cao, Zizheng and Tang, Ming and Liu, Deming and Giles, Ian and Koonen, Ton and Okonkwo, Chigo},
	month = may,
	year = {2016},
	pages = {2302},
}

@misc{becerra-deana_fabrication_2024,
	title = {Fabrication and {Characterization} of {Photonic} {Lanterns} {Using} {Coupled}-{Mode} {Theory}},
	url = {http://arxiv.org/abs/2411.02182},
	doi = {10.48550/arXiv.2411.02182},
	language = {en},
	urldate = {2025-03-03},
	publisher = {arXiv},
	author = {Becerra-Deana, Rodrigo Itzamná and Ramadier, Guillaume and Sivry-Houle, Martin Poinsinet de and Maltais-Tariant, Raphael and Virally, Stéphane and Boudoux, Caroline and Godbout, Nicolas},
	month = nov,
	year = {2024},
	note = {arXiv:2411.02182 [physics]},
}

@article{zia_interferometric_2023,
	title = {Interferometric imaging of amplitude and phase of spatial biphoton states},
	volume = {17},
	issn = {1749-4885, 1749-4893},
	url = {https://www.nature.com/articles/s41566-023-01272-3},
	doi = {10.1038/s41566-023-01272-3},
	language = {en},
	number = {11},
	urldate = {2024-08-23},
	journal = {Nature Photonics},
	author = {Zia, Danilo and Dehghan, Nazanin and D’Errico, Alessio and Sciarrino, Fabio and Karimi, Ebrahim},
	month = nov,
	year = {2023},
	pages = {1009--1016},
}

@article{xin_laboratory_2024,
	title = {Laboratory demonstration of a {Photonic} {Lantern} {Nuller} in monochromatic and broadband light},
	volume = {10},
	issn = {2329-4124},
	url = {https://www.spiedigitallibrary.org/journals/Journal-of-Astronomical-Telescopes-Instruments-and-Systems/volume-10/issue-02/025001/Laboratory-demonstration-of-a-Photonic-Lantern-Nuller-in-monochromatic-and/10.1117/1.JATIS.10.2.025001.full},
	doi = {10.1117/1.JATIS.10.2.025001},
	language = {en},
	number = {02},
	urldate = {2025-02-28},
	journal = {Journal of Astronomical Telescopes, Instruments, and Systems},
	author = {Xin, Yinzi and Echeverri, Daniel and Jovanovic, Nemanja and Mawet, Dimitri and Leon-Saval, Sergio and Amezcua-Correa, Rodrigo and Yerolatsitis, Stephanos and Fitzgerald, Michael P. and Gatkine, Pradip and Kim, Yoo Jung and Lin, Jonathan and Norris, Barnaby and Ruane, Garreth and Sallum, Steph},
	month = apr,
	year = {2024},
}

@inproceedings{eikenberry_photonic_2024,
	address = {Yokohama, Japan},
	title = {Photonic quantum-inspired sub-diffraction imager},
	isbn = {978-1-5106-7515-5 978-1-5106-7516-2},
	url = {https://www.spiedigitallibrary.org/conference-proceedings-of-spie/13096/3019152/Photonic-quantum-inspired-sub-diffraction-imager/10.1117/12.3019152.full},
	doi = {10.1117/12.3019152},
	language = {en},
	urldate = {2024-11-04},
	booktitle = {Ground-based and {Airborne} {Instrumentation} for {Astronomy} {X}},
	publisher = {SPIE},
	author = {Eikenberry, Stephen S. and Romer, Miguel and Crowe, Tara and Conwell, Robert and Dobias, Caleb and Batarseh, Ameer and Moraitis, Christina D. and Cruz-Delgado, Daniel and Yerolatsitis, Stephanos and Amezcua-Correa, Rodrigo and Bandres, Miguel A. and Leon-Saval, Sergio G. and Thibaut, Sarah and Miller, Vincent and Akers, Aiden and Donaldson-Hanna, Kerri and Cooper, Matthew},
	editor = {Vernet, Joël R. and Bryant, Julia J. and Motohara, Kentaro},
	month = jul,
	year = {2024},
	pages = {26},
}

@article{marcuse_loss_1977,
	title = {Loss {Analysis} of {Single}-{Mode} {Fiber} {Splices}},
	volume = {56},
	issn = {00058580},
	url = {https://ieeexplore.ieee.org/document/6768444},
	doi = {10.1002/j.1538-7305.1977.tb00534.x},
	language = {en},
	number = {5},
	urldate = {2024-04-09},
	journal = {Bell System Technical Journal},
	author = {Marcuse, D.},
	month = may,
	year = {1977},
	pages = {703--718},
}

@article{choudhury_computational_2020,
	title = {Computational optical imaging with a photonic lantern},
	volume = {11},
	issn = {2041-1723},
	url = {https://www.nature.com/articles/s41467-020-18818-6},
	doi = {10.1038/s41467-020-18818-6},
	language = {en},
	number = {1},
	urldate = {2024-05-06},
	journal = {Nature Communications},
	author = {Choudhury, Debaditya and McNicholl, Duncan K. and Repetti, Audrey and Gris-Sánchez, Itandehui and Li, Shuhui and Phillips, David B. and Whyte, Graeme and Birks, Tim A. and Wiaux, Yves and Thomson, Robert R.},
	month = oct,
	year = {2020},
	pages = {5217},
}

@article{norris_optimal_2022,
	title = {Optimal broadband starlight injection into a single-mode fibre with integrated photonic wavefront sensing},
	volume = {30},
	issn = {1094-4087},
	url = {https://opg.optica.org/abstract.cfm?URI=oe-30-19-34908},
	doi = {10.1364/OE.465639},
	language = {en},
	number = {19},
	journal = {Optics Express},
	author = {Norris, Barnaby and Betters, Christopher and Wei, Jin and Yerolatsitis, Stephanos and Amezcua-Correa, Rodrigo and Leon-Saval, Sergio},
	month = sep,
	year = {2022},
	pages = {34908},
}

@article{norris_all-photonic_2020,
	title = {An all-photonic focal-plane wavefront sensor},
	volume = {11},
	issn = {2041-1723},
	url = {https://www.nature.com/articles/s41467-020-19117-w},
	doi = {10.1038/s41467-020-19117-w},
	language = {en},
	number = {1},
	journal = {Nature Communications},
	author = {Norris, Barnaby R. M. and Wei, Jin and Betters, Christopher H. and Wong, Alison and Leon-Saval, Sergio G.},
	month = oct,
	year = {2020},
	pages = {5335},
}

@inproceedings{gallagher_eigenmode_2003,
	address = {San Jose, CA},
	title = {Eigenmode expansion methods for simulation of optical propagation in photonics: pros and cons},
	shorttitle = {Eigenmode expansion methods for simulation of optical propagation in photonics},
	url = {http://proceedings.spiedigitallibrary.org/proceeding.aspx?doi=10.1117/12.473173},
	doi = {10.1117/12.473173},
	language = {en},
	booktitle = {{SPIE} {Proceedings}},
	publisher = {SPIE},
	author = {Gallagher, Dominic F. G. and Felici, Thomas P.},
	editor = {Sidorin, Yakov S. and Tervonen, Ari},
	month = jun,
	year = {2003},
	note = {ISSN: 0277-786X},
}

@article{velazquez-benitez_scaling_2018,
	title = {Scaling photonic lanterns for space-division multiplexing},
	volume = {8},
	copyright = {https://creativecommons.org/licenses/by/4.0},
	issn = {2045-2322},
	url = {https://www.nature.com/articles/s41598-018-27072-2},
	doi = {10.1038/s41598-018-27072-2},
	language = {en},
	number = {1},
	journal = {Scientific Reports},
	author = {Velázquez-Benítez, Amado M. and Antonio-López, J. Enrique and Alvarado-Zacarías, Juan C. and Fontaine, Nicolas K. and Ryf, Roland and Chen, Haoshuo and Hernández-Cordero, Juan and Sillard, Pierre and Okonkwo, Chigo and Leon-Saval, Sergio G. and Amezcua-Correa, Rodrigo},
	month = jun,
	year = {2018},
	note = {Publisher: Springer Science and Business Media LLC},
}

@inproceedings{romer_broadband_2025,
	address = {San Francisco, United States},
	title = {Broadband photonic lantern transfer matrix characterization for wavefront sensing},
	isbn = {978-1-5106-8494-2 978-1-5106-8495-9},
	url = {https://www.spiedigitallibrary.org/conference-proceedings-of-spie/13373/3043722/Broadband-photonic-lantern-transfer-matrix-characterization-for-wavefront-sensing/10.1117/12.3043722.full},
	doi = {10.1117/12.3043722},
	language = {en},
	booktitle = {Photonic {Instrumentation} {Engineering} {XII}},
	publisher = {SPIE},
	author = {Romer, Miguel A. and Batarseh, Ameer B. and Crowe, Tara and Conwell, Robert and Dobias, Caleb and Cruz-Delgado, Daniel and Bandres, Miguel A. and Amezcua-Correa, Rodrigo and Eikenberry, Stephen S.},
	editor = {Soskind, Yakov and Busse, Lynda E.},
	month = mar,
	year = {2025},
	pages = {26},
}

@misc{chandrasekharan_high-throughput_2025,
	title = {High-throughput polarization-independent spatial mode shaping with mode-selective photonic lanterns},
	url = {http://arxiv.org/abs/2506.08595},
	doi = {10.48550/arXiv.2506.08595},
	language = {en},
	publisher = {arXiv},
	author = {Chandrasekharan, Harikumar K. and Donaldson, Ross},
	month = jun,
	year = {2025},
	note = {arXiv:2506.08595 [physics]},
}

@inproceedings{milne_coherent_2023,
	address = {Munich, Germany},
	title = {Coherent {Beam} {Shaping} with {Multicore} {Fiber} {Photonic} {Lanterns}},
	copyright = {https://doi.org/10.15223/policy-029},
	isbn = {9798350345995},
	url = {https://ieeexplore.ieee.org/document/10232791/},
	doi = {10.1109/CLEO/Europe-EQEC57999.2023.10232791},
	language = {en},
	booktitle = {2023 {Conference} on {Lasers} and {Electro}-{Optics} {Europe} \& {European} {Quantum} {Electronics} {Conference} ({CLEO}/{Europe}-{EQEC})},
	publisher = {IEEE},
	author = {Milne, A. and Parker, H. E. and Wright, T. A. and Benoît, A. and Harrington, K. and Leach, J. and Phillips, D. B. and Stone, J. M. and Birks, T. A. and Thomson, R. R.},
	month = jun,
	year = {2023},
	pages = {1--1},
}

@article{carpenter_all_2012,
	title = {All {Optical} {Mode}-{Multiplexing} {Using} {Holography} and {Multimode} {Fiber} {Couplers}},
	volume = {30},
	issn = {0733-8724, 1558-2213},
	url = {http://ieeexplore.ieee.org/document/6177203/},
	doi = {10.1109/JLT.2012.2191586},
	language = {en},
	number = {12},
	journal = {Journal of Lightwave Technology},
	author = {Carpenter, Joel and Wilkinson, Timothy D.},
	month = jun,
	year = {2012},
	pages = {1978--1984},
}

@inproceedings{fontaine_hermite-gaussian_2021,
	address = {Washington, DC},
	title = {Hermite-{Gaussian} mode multiplexer supporting 1035 modes},
	isbn = {978-1-943580-86-6},
	url = {https://opg.optica.org/abstract.cfm?URI=OFC-2021-M3D.4},
	doi = {10.1364/OFC.2021.M3D.4},
	language = {en},
	booktitle = {Optical {Fiber} {Communication} {Conference} ({OFC}) 2021},
	publisher = {Optica Publishing Group},
	author = {Fontaine, Nicolas K. and Chen, Haoshuo and Mazur, Mikael and Dallachiesa, Lauren and Kim, K.W. and Ryf, Roland and Neilson, David and Carpenter, Joel},
	year = {2021},
	pages = {M3D.4},
}

@article{jovanovic_2023_2023,
	title = {2023 {Astrophotonics} {Roadmap}: pathways to realizing multi-functional integrated astrophotonic instruments},
	volume = {5},
	issn = {2515-7647},
	shorttitle = {2023 {Astrophotonics} {Roadmap}},
	url = {https://iopscience.iop.org/article/10.1088/2515-7647/ace869},
	doi = {10.1088/2515-7647/ace869},
	language = {en},
	number = {4},
	journal = {Journal of Physics: Photonics},
	author = {Jovanovic, Nemanja and Gatkine, Pradip and Anugu, Narsireddy and Amezcua-Correa, Rodrigo and Basu Thakur, Ritoban and Beichman, Charles and Bender, Chad F. and Berger, Jean-Philippe and Bigioli, Azzurra and Bland-Hawthorn, Joss and Bourdarot, Guillaume and Bradford, Charles M and Broeke, Ronald and Bryant, Julia and Bundy, Kevin and Cheriton, Ross and Cvetojevic, Nick and Diab, Momen and Diddams, Scott A and Dinkelaker, Aline N and Duis, Jeroen and Eikenberry, Stephen and Ellis, Simon and Endo, Akira and Figer, Donald F and Fitzgerald, Michael P. and Gris-Sanchez, Itandehui and Gross, Simon and Grossard, Ludovic and Guyon, Olivier and Haffert, Sebastiaan Y and Halverson, Samuel and Harris, Robert J and He, Jinping and Herr, Tobias and Hottinger, Philipp and Huby, Elsa and Ireland, Michael and Jenson-Clem, Rebecca and Jewell, Jeffrey and Jocou, Laurent and Kraus, Stefan and Labadie, Lucas and Lacour, Sylvestre and Laugier, Romain and Ławniczuk, Katarzyna and Lin, Jonathan and Leifer, Stephanie and Leon-Saval, Sergio and Martin, Guillermo and Martinache, Frantz and Martinod, Marc-Antoine and Mazin, Benjamin A and Minardi, Stefano and Monnier, John D and Moreira, Reinan and Mourard, Denis and Nayak, Abani Shankar and Norris, Barnaby and Obrzud, Ewelina and Perraut, Karine and Reynaud, François and Sallum, Steph and Schiminovich, David and Schwab, Christian and Serbayn, Eugene and Soliman, Sherif and Stoll, Andreas and Tang, Liang and Tuthill, Peter and Vahala, Kerry and Vasisht, Gautam and Veilleux, Sylvain and Walter, Alexander B and Wollack, Edward J and Xin, Yinzi and Yang, Zongyin and Yerolatsitis, Stephanos and Zhang, Yang and Zou, Chang-Ling},
	month = oct,
	year = {2023},
	pages = {042501},
}

@article{lin_real-time_2023,
	title = {Real-time {Experimental} {Demonstrations} of a {Photonic} {Lantern} {Wave}-front {Sensor}},
	volume = {959},
	issn = {2041-8205, 2041-8213},
	url = {https://iopscience.iop.org/article/10.3847/2041-8213/ad12a4},
	doi = {10.3847/2041-8213/ad12a4},
	language = {en},
	number = {2},
	journal = {The Astrophysical Journal Letters},
	author = {Lin, Jonathan W. and Fitzgerald, Michael P. and Xin, Yinzi and Kim, Yoo Jung and Guyon, Olivier and Norris, Barnaby and Betters, Christopher and Leon-Saval, Sergio and Ahn, Kyohoon and Deo, Vincent and Lozi, Julien and Vievard, Sébastien and Levinstein, Daniel and Sallum, Steph and Jovanovic, Nemanja},
	month = dec,
	year = {2023},
	pages = {L34},
}

@article{birks_photonic_2015,
	title = {The photonic lantern},
	volume = {7},
	issn = {1943-8206},
	url = {https://opg.optica.org/abstract.cfm?URI=aop-7-2-107},
	doi = {10.1364/AOP.7.000107},
	language = {en},
	number = {2},
	journal = {Advances in Optics and Photonics},
	author = {Birks, T. A. and Gris-Sánchez, I. and Yerolatsitis, S. and Leon-Saval, S. G. and Thomson, R. R.},
	month = jun,
	year = {2015},
	pages = {107},
}

@article{thyng2016true,
  title={True colors of oceanography: Guidelines for effective and accurate colormap selection},
  author={Thyng, Kristen M and Greene, Chad A and Hetland, Robert D and Zimmerle, Heather M and DiMarco, Steven F},
  journal={Oceanography},
  volume={29},
  number={3},
  pages={9--13},
  year={2016},
  publisher={JSTOR}
}

@article{gabor1948new,
  title={A new microscopic principle.},
  author={Gabor, Dennis},
  year={1948},
  journal={Nature}
}

@article{canny2009computational,
  title={A computational approach to edge detection},
  author={Canny, John},
  journal={IEEE Transactions on pattern analysis and machine intelligence},
  number={6},
  pages={679--698},
  year={2009},
  publisher={Ieee}
}



\section*{Supplementary material}

# Illuminating the lantern: coherent, spectro-polarimetric characterisation of a multimode converter (supplemental document)

## 1. METHODS: ADDITIONAL INFORMATION

Figure S1 illustrates a system level view of the setup. Light from a laser or superluminescent diode (SLD) is first split for wavelength measurement (laser only), then split into two beams: one for injection (bottom row) and one for the reference beam (middle row). Each arm is them combined on the detector to induce interference for the inference of the complex electric field. Switching optics such as the shutter, fibre injection stage and polarisation switch are operated remotely to improve the thermal stability of the system.

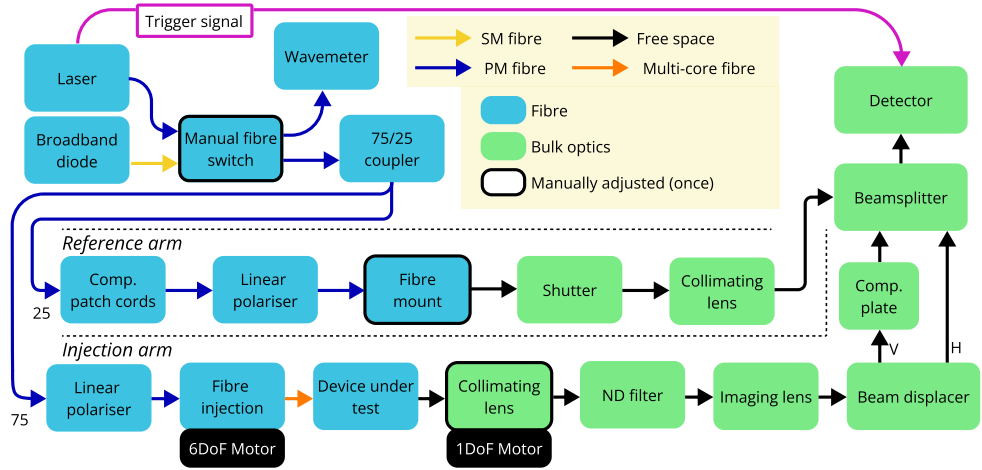

**Fig. S1. System block diagram for digital off-axis holography.** Fibre and free space optical components are shown in blue and green respectively. A 6 degrees of freedom (DoF) motor injects light into a single port, while a 1 DoF linear stage changes the path length of one arm with respect to the other. The bottom row follows the device injection arm, and the one above follows the reference arm. Coarse path length matching is achieved with compensation ("Comp.") patch cords and glass plates. As the system requires stability for many measurements, all of the switching optics are operated remotely, with only some optics adjusted manually once for focus.

We choose to capture frames at a high duty cycle during a continuous wavelength sweep for several reasons. Firstly, the IMX990 InGaAs detector used in the setup showed signs of internal fringing that was dependent on the wavelength (with sharp responses at a sub nanometre level), pixel position and angle of incidence on the light. This is intractable to calibrate, and hence we move to a shorter coherence length regime, where such behaviour is smoothed out during an exposure. An equivalent setup could use a monochromator. In sweeping over wavelengths continuously, we are also able to sample a wide bandwidth over a very short ($< 1\,$s) window, reducing the effects of phase drifts from the stability of the room. Finally, the wavelength sweep captures data at the white light fringe, hence probing the dispersion of the device under test.

**Wavelength calibration**

Figure S2 illustrates the two pieces of information needed to measure which frame corresponds to which wavelength. First, we measure the functional form of the wavelength sweep $\lambda(t)$ using a wavemeter, and then we monitor a wavelength dependent metric (such as the phase of the off-axis component) using the precise camera timestamps. Combined these give a mapping from frame index to wavelength.

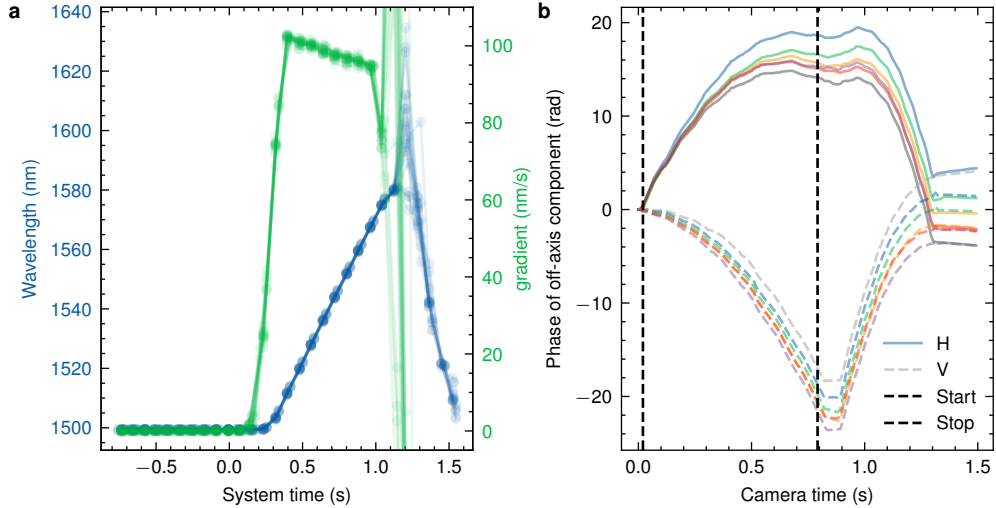

**Fig. S2. Wavelength to frame calibration. a** Wavelength (blue) and its gradient (green) captured using a wavemeter as a function of time relative to issuing the start of the sweep command, sampled on the machine that issues commands. The sweep itself is weakly non-linear, with the gradient changing from just over 100 nm/s to 90 nm/s. **b** Phase of the off-axis holographic component (central pixels of the white box in Figure 1) depends on the wavelength, and hence serves as a good metric to find where the sweep starts and ends (clearest for V polarisation). We mark the first and last frame with black dashed lines.

**Exciting a single principal mode**

To excite a single principal mode with minimal crosstalk, we must operate the fibre alignment stage at a distance that has an effective mode field diameter of less than $10\mu$m. The separation between cores is $60\mu$m, and hence the set of motor states that excites a single port clearly is sparse. We find this set in two stages, summarised in each row of Figure S3. First we add $\sim 130\,\mu$m of distance between the fibres to widen the regions in $(x, y)$ where injection occurs, enabling a coarser sampling to take place, decreasing the time taken. We do a raster scan, measuring the total flux on the camera at each point, shown in Figure S3 **a**. The flux is then fit as the sum of Gaussian functions, with centres lying on a finite hexagonal grid. We fit the origin, separation and rotation of the grid, as well as the nuisance parameters of the mode field diameter and the individual amplitude. There is no need to fit an offset term as all images taken are dark subtracted; images with no coupling have an expected total flux value of zero. We then take a step closer to the final distance and visit each port, sampling point around it and fitting the overall grid again. This is repeated until we are close enough such that the width of the Gaussian around each peak is around 1/5 the port separation, sufficiently rejecting the light in other modes. We then validate the setup by visiting each port, refining the positions and estimate of the grid, as well as adjusting the laser power used on the whole set to avoid saturation. To visit each port, we conduct a local search. This involves sampling points drawn from a truncated Gaussian distribution centred on an initial guess, then fitting a Gaussian to the flux measured at each position (shown in Figure S3 **d**-**f**). This is done twice – first with a wider Gaussian distribution to get a coarse estimate, and then refined using both previous samples and more samples from a narrower distribution. The number of samples was tuned empirically and robust behaviour was found with 20 coarse and 20 fine samples. The hexagonal grid is refined further using these measurements.

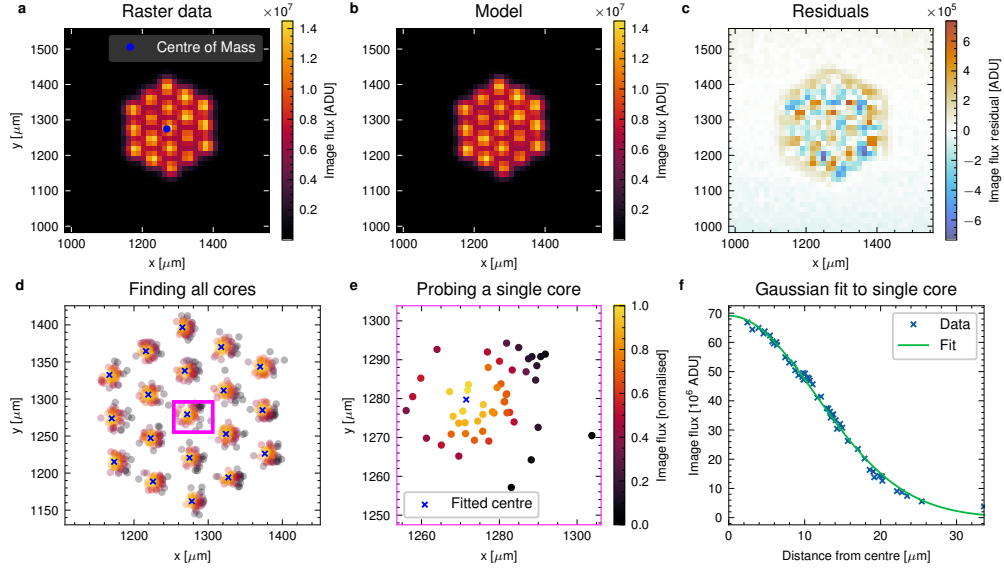

**Fig. S3. Optimization of injection of light from a single-mode fibre into a single port of a multicore fibre. a** Measuring total flux through a raster of $(x, y)$ positions, taken after pulling away from the fibre face. **b** Fit from a sum of Gaussians model, where the centres lie on a finite hexagonal grid. **c** Map of residuals between the data and model, note the difference in the order of magnitude on the colorbar. **d** After approach to the final distance, we sample points near ports. **e** Zoom of sampling near a single port. **f** Normalised fluxes for a single port as a function of distance from the fitted centre. Each port is well fit by a Gaussian model.

**Digital off-axis holography parameter fitting**

As outlined in the main text, we need to fit the centres $(x(\lambda), \theta(\lambda))$ and widths of features in both the image and Fourier spaces to correctly reconstruct the complex field. For the image space parameters, we leverage the insight that the sum over all mode fields supported by the lantern gives a near top-hat function, effectively re-imaging the core on the detector. This is shown in panel **a** of Figure S4. A rough approximation of the centre (to within less than half the radius) can be found with the centre of mass, however asymmetries within the modes can bias this measurement away from the geometrical centre. This is problematic when computing overlap integrals later in the pipeline. Hence, to refine the estimate of the centre in the image space, we use the Canny edge detection [1] algorithm and then estimate the centre using a least squares circle estimator [2]. This uses geometric features and hence is more robust to asymmetries in resulting top hat function.

We must consider the chromaticity of the optics in the beam train when fitting these parameters. From refractive index tables, the calcite beam displacer alone causes a shift of around 2 pixels over a 50 nm. We account for this by fitting a degree 2 polynomial in $\lambda$ to each of the $x$ and $y$ coordinates at each output polarisation, shown in panel **d** of Figure S4.

The above process also trivially gives the radius of the circular feature, which is then used to determine the size of the window before taking a Fourier transform.

We now turn to the fitting of parameters in Fourier space. The structure follows similarly to the image space, where we must find the centre $(\theta(\lambda))$ and radius of the off-axis component of the power spectrum.

To find the geometric centre with sub-pixel precision, we solve the optimisation problem

$$\arg\min_{x} (r(x)),  \quad (S1)$$

$$\text{s.t.} \ \sum \left( C_x(r(x)) \hat{P} \right) = \rho, \quad (S2)$$

where $C_x(r(x))$ is a soft edged circle with centre $x = (x_c, y_c)$ and radius $r(x)$, which is defined implicitly as the radius at which the circle encloses a fraction $\rho \in (0, 1)$ of the normalised power spectrum $\hat{P}$. We also smooth the power spectrum with a Gaussian filter beforehand to improve convergence. Intuitively, the above finds the smallest possible circle that encloses a

3

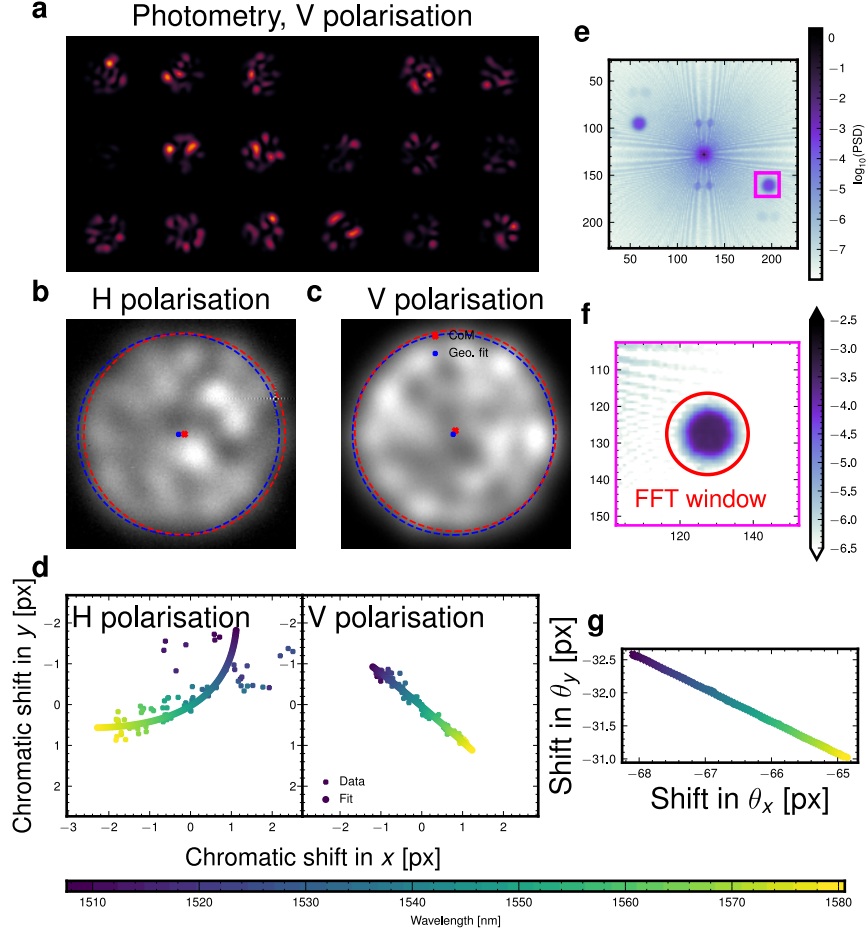

**Fig. S4. Parameter fitting for digital off-axis holography.** All data in the first two rows is at a single wavelength. The left column shows image space fitting: **a** photometry frames from 18 ports, **b,c** co-added photometry with estimated centres, **d** fitting centres as a function of wavelength. The right column shows the same process in Fourier space: **e** off-axis term in the power spectrum from co-adding, **f** sub-pixel centring and **h** fitting through wavelength.

significant fraction of the energy in the image, identifying the boundary the off-axis feature and the background. We use the ∂Lux [3] implementation of a soft circle and the solver in Optimstix [? ] and Jax [4], solving a root finding problem within a minimisation problem. The fraction $\rho$ is selected empirically to be 0.9 which gives a good balance of robustness to asymmetry in the circle and noise outside the circle.

For each frame taken with the shutter open (over all polarisations), we incoherently sum the power spectra to build a single high signal to noise ratio (SNR) power spectrum, plotted on a logarithm strech in panel **e** of Figure S4. Next, we use the fact that the off-axis component is half the radius of the central component to initialise a off-axis centre. The scale of the central peak is given by the first point of inflection in the encircled energy as a function of radius. This region is then masked and the resulting power spectrum is convolved with a Gaussian with the expected scale. A simple maximum then yields the centre of the off-axis component to within a few pixels, enabling us to apply the above optimisation procedure in Equation S1 to fine tune the estimate. Again, we consider that this centre must shift with wavelength, however this is much more constrained than in the image plane: the motion must be in a line, with spacing in wavelength obeying

$$\boldsymbol{\theta}(\lambda) = \frac{\boldsymbol{\theta}_0 \lambda_0}{\lambda}, \tag{S3}$$

where the zero subscript refers to some arbitrarily chosen zero point. We find that the line fit

4

doesn't precisely go through the origin, and instead fit this as an additional parameter. This could be eliminated through the use of reflective instead of refractive optics to collimate the reference arm. The centres of these off-axis components are illustrated in panel **g** of Figure S4.

**Manufacturing process**

The photonic lantern used in this work was fabricated from a 19-core multicore fibre (MCF) produced by co-author SLS at the University of Bath. The MCF has 6.5 $\mu$m cores with numerical aperture 0.14, arranged on a 60 $\mu$m hexagonal pitch within a 320 $\mu$m cladding. The low-index jacket capillary that forms the multimode output cladding is from Photonics Bretagne (PB). The base outer diameter (OD) and inner diameter (ID) are 1760 $\mu$m and 750 $\mu$m, respectively.

The multicore fibre is inserted into a pre-tapered piece of the PB capillary, whose inner diameter is slightly larger than the MCF, and the assembly is tapered on a 3SAE Combined Manufacturing Station (CMS) to create a 6 cm adiabatic transition. At the multimode end, the effective core diameter is 34.7 $\mu$m with numerical aperture 0.126. After tapering, the device is terminated in a standard SMA connector, and a 3D-printed insert protects the tapered section during handling and integration.

**Simulation**

The results for the simulated device in the main text are generated using `Rsoft Beamprop`. A model of the 19-core PL was created using a linear taper, with core size and separation, taper ratio and length, and refractive indexes based on known values for the physical device. Light was launched into single-mode ports one at a time and propagated to the multimode end. The resulting mode field was decomposed into LP modes via an overlap integral and their complex coefficients recorded. This was repeated for 10 wavelengths spread evenly between 1507.5 nm and 1580.5 nm.

## 2. COMPONENT LIST

Table S1 lists the major components used. For a complete list including cables, posts etc. and any associated computer aided design (CAD) files, contact the corresponding author.

**Table S1.** Components in the characterisation setup, listed in order of propagation.

| Description | Supplier | Part number | Notes |
|---|---|---|---|
| *Pre-split components* | | | |
| Swept wavelength source | Anritsu | MG9637A | |
| Broadband SLD | Thorlabs | SLD1005S | Driven by CLD1015 |
| Coupler | Thorlabs | PW1550R3F1 | |
| *Injection beam* | | | |
| Inline fibre polariser | AFW | ILP-15-L-1-5-1-SA | |
| Fibre alignment stage | Luminos | I6000 | Custom interface plates |
| Delay line stage | Newport | UTS150CC | Driven by ESP300 controller |
| Collimating lens | Thorlabs | C20SMA-C | |
| ND filter | Thorlabs | NENIR20A-C | |
| Re-imaging lens | Thorlabs | AC254-400-C-ML | |
| Beam displacer | Thorlabs | BD27 | |
| Compensating plate | Custom | | Microscope slides, NOA61 |
| *Reference beam* | | | |
| Inline fibre polariser | AFW | ILP-15-C-1-1-11-SA | |
| Shutter | Thorlabs | SH05/M | |
| Collimating lens | Thorlabs | AC254-200-C-ML | |
| *Beam recombination* | | | |
| Beamsplitter | Thorlabs | CCM1-BS015/M | |
| Detector | Lucid | Atlas SWIR 1.3MP IP67 | |
| *Electronics* | | | |
| Trigger controller | Arduino | Nano | |

## 3. RESULTS: ADDITIONAL INFORMATION

**Validation with a single-mode fibre**

We first evaluate the performance of our system on a known, simple device: a SMF. This is shown in Figure S5. The electric fields reconstructed from the SMF appear near uniform in phase and Gaussian in amplitude. The dominant term in the phase residual is a tilt with a small amount of defocus. For the SMF this could be optimized further, however gives a realistic impression of how finely we can adjust the focus in the photonic lantern case where the field is not known to be uniform in phase. The differential tilt seen in Figure S5 **b** is consistent with wedge tolerances on the polarisation beam displacer and compensation plate, which lie in an intermediate plane.

When considering the behaviour of the mode field as a function of wavelength, we observe oscillations in both the phase and amplitude of the complex coefficients. This is consistent with the fringe power in delay space, where we observe a secondary peak of relative height $A_r = 0.25$ at a distance of $d = 192\,\mu$m (as is the case for V, using the two peaks around 29.5 mm in Figure S5 **d** only). This secondary peak contributes another phasor to the fundamental signal, which rotates faster in the complex plane than the fundamental due to the different optical path difference.

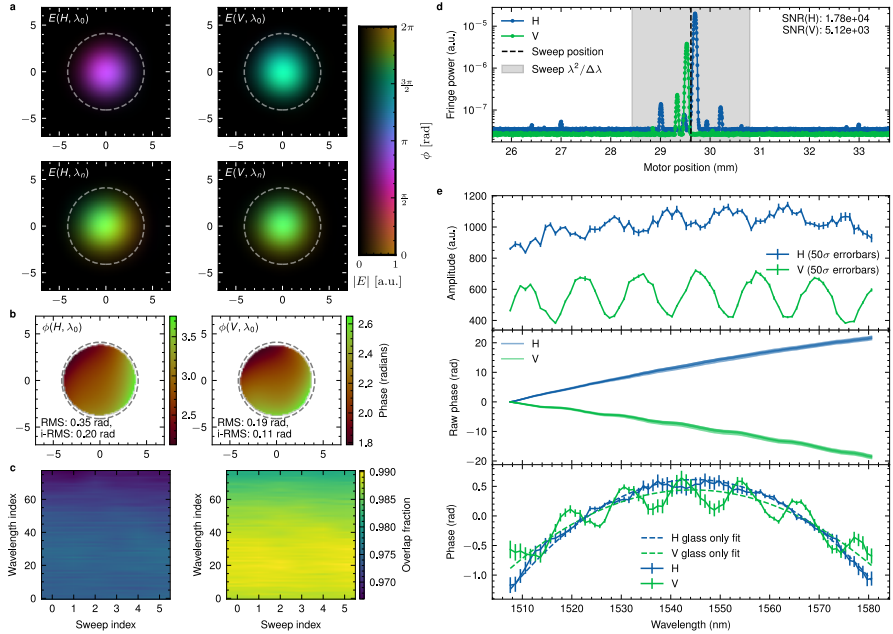

**Fig. S5. Reconstruction validation on a single-mode fibre (SMF), with monochromatic performance (left) and wavelength-coherent reconstruction (right). a** Reconstructed electric field for both polarisations at two different wavelengths show phase uniformity and a Gaussian structure. All image plane axes units are in $\mu$m. **b** The phase over the face of the fibre is flat to within 0.20 rad rms, weighted by intensity. **c** The overlap integral explainability shows that the reconstructed field is at worst 97% explained by a single LP01 mode, as expected. **d** Probing the position of the white light fringe by measuring the fringe power as a function of delay with a broadband source. The grey shaded region shows the effective coherence length during the laser sweep, and the dashed black line shows the position of the delay line motor during the sweep. The fringe visibility using a broadband source shows clear peaks and validates the phase unwraping used in the following step. Secondary peaks are likely due to birefringence in fibre optics. **e** Complex LP01 modal coefficients as a function of wavelength. The average amplitude over six sweeps is very consistent – error bars show $50\sigma$ (standard deviation). Raw phases (middle row) of the reconstructed single-mode are dominated by linear trends induced from path difference between polarisations. Once removed (bottom row, $1\sigma$ error bars), the dispersion properties are consistent with a common differential air path of -165.3 mm and a glass path of 141.4 mm or 144.8 mm for H and V respectively, plus oscillations consistent with birefringence as evident in **d** (see main text for details).

Hence, the coherent light in from the secondary peak induces oscillations in amplitude with a factor $A_r = 0.25$ or phase of $\arctan(A_r) \approx 0.25$ rad, both with a period given by $\lambda^2/d = 11.25$ nm, corresponding to between 6 and 7 oscillations over a 75 nm bandpass. This is all very consistent with the data shown in the top and bottom panels of Figure S5 **e**. These oscillations are similarly present in H polarisation, however are much smaller in amplitude as the main peak has higher power. We hypothesise that the peaks are from birefringence, likely from tolerances on the connectors between birefringent polarisation maintaining fibres.

We also can compare this to the phase evolution shown in the main text. For the photonic lantern, the quadratic coefficient in phase is stronger than the SMF as about 116 mm more compensating fibre in the injection arm was required to find the white light fringe, despite the length of the photonic lantern with multicore fibre being close to that of the SMF used in validation. This indicates that the dispersion of the multicore fibre (MCF) is different to the SMF.

### Broadband coherence properties

Next, we turn to the coherence properties of the lantern, illustrated in Figure S6. Each row is formed from taking a single port from the outer edges of the hexagonal grid, plus the central core.

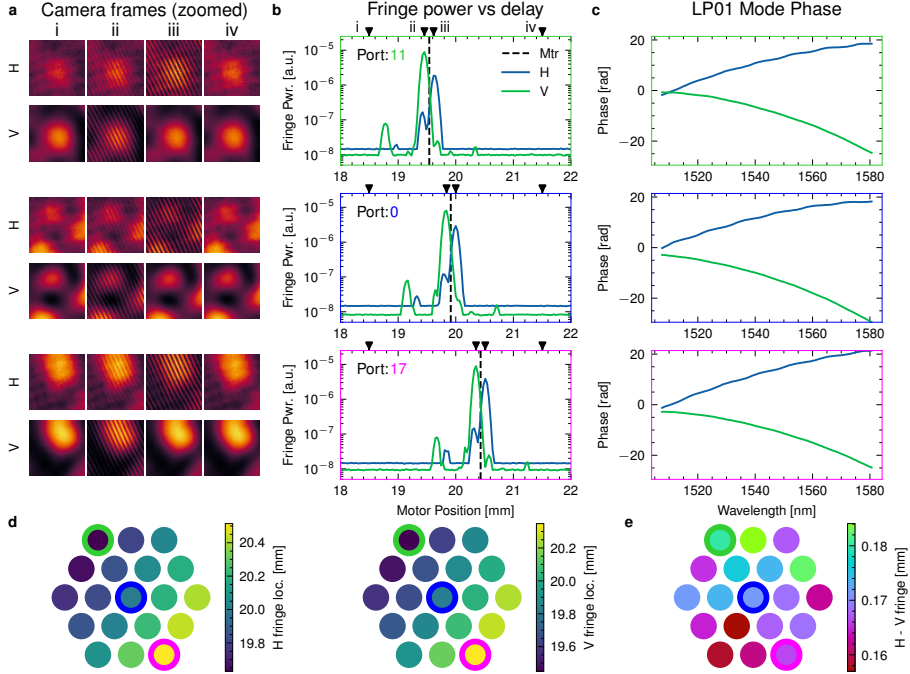

**Fig. S6.** Characterisation of the broadband coherence properties of the MCF fed photonic lantern. **a** Zoomed images of both polarisations in broadband light for different delay line motor positions for 3 different ports in both polarisations. **b** The fringe power as a function of motor position identifies the white light fringe. Arrows above plots indicate the positions of the frames shown on the left, with the maximum fringe power position shown in ii for V polarisation and iii for H polarisation. The final motor position for data capture is indicated with a black dashed line. **c** The reconstructed phase of the LP01 mode as a function of wavelength. This is consistent with the broadband results, with oscillations more apparent in the case of stronger secondary peaks in fringe power, and a larger linear trend when data is taken further from the white light fringe as in port 5 vs. others. **d** Visualisation of the peak position over all ports, corresponding to the physical position of ports in the MCF. We find a trend over the fringe location consistent with a bend dependent change in the refractive index of the multicore fibre that feeds the photonic lantern (see text for details). Coloured edges around circles correspond to the ports highlighted in other rows. **e** The difference between fringe locations is relatively constant, but with variations at a level slightly larger than the delay line step size (32 $\mu$m).

This is visualised in the bottom row, where the circle edge colour matches the axis colour. The fringe power metric finds the white light fringe, located at ii and iii for V and H respectively. All white light fringes are found with a SNR greater than $10^3$ using a median and median absolute deviation analysis. This metric is selected to avoid bias towards secondary peaks (which are part of the signal we are interested in). The picture in this delay space demonstrates a clear shift of about 750 $\mu$m in the position of the white light fringe along a direction on the multicore fibre. This is much larger than any tip/tilt induced effects at the fibre stage and hence shows a true differential delay. We believe that such a large effect is likely due to a bend effect in the 1 m of multicore fibre rather than something in the photonic lantern itself. Again, secondary peaks are present in delay space, however, the peaks are at different relative heights between ports. This indicates that the birefringence depends on the device, and cannot simply be calibrated out relative to a single-mode fibre. We also see qualitatively the same behaviour, where the polarisation with a relatively larger secondary peak (H in this case, was V previously) has more visible oscillations in the phase of the modes when using the swept wavelength source. We also analyse the difference in the peak positions for each polarisation, using a cubic spline interpolation to sub-sample the motor steps (32 $\mu$m) and estimate the peak location. Differences in white light position between polarisations are weakly present, with a maximum difference of 40 $\mu$m and no

dominant direction in the physical units of the hexagonal grid.

**Throughput and explainability**

An important metric we monitor is the "explainability" of the field in the modal basis we project into, defined as $\sqrt{\sum_j |\eta_{i,j}|}$. This represents the fraction (in an amplitude sense) of the field that is consistent with basis modes alone, with values close to unity indicating a good match. The square of this value is the total power explained by the basis.

Figure S7 presents the throughput and LP mode explainability per port and wavelength. There is no significant trend in the total throughput, but we include it for completeness. The LP mode explainability, on the other hand, shows a high agreement, with median value of 0.937 and 0.949 for H and V over all ports and wavelengths. Lower values seen in oscillations for port 7 in H are indicative of low flux. Visualising this in the hexagonal grid of the multicore fibre (bottom row) reveals a correlation between ports, and a weak gradient where the ports that are physically lower have higher LP mode overlap.

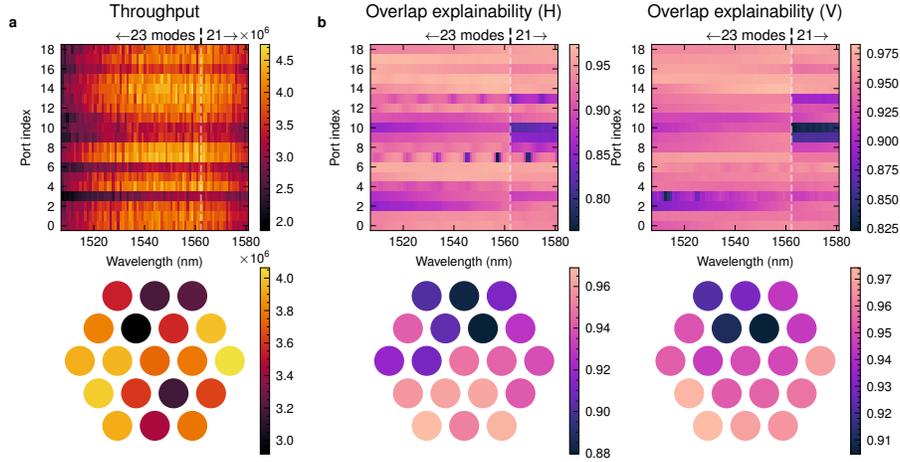

**Fig. S7. Relative throughput and LP mode explainability of fields. a** Throughput for each port and wavelength, measured by photometry. We also illustrate the median value on a hexagonal grid. There is no obvious structure in either case. **b** The overlap integral $\eta_{i,j}$ over all ports and wavelengths, and plotted on a hexagonal grid. The white dashed line indicates the wavelength at which a multimode fibre with a diameter equal to that of the device begins to guide 21 instead of 23 modes. For some fields there is a noticeable drop in the explainability of the field as a superposition of LP modes, however for others this change has little to no effect.

## REFERENCES

1. J. Canny, "A computational approach to edge detection," IEEE Trans. on pattern analysis machine intelligence pp. 679–698 (2009).
2. C. F. Jekel, "Obtaining non-linear orthotropic material models for pvc-coated polyester via inverse bubble inflation," Ph.D. thesis, Stellenbosch: Stellenbosch University (2016).
3. L. Desdoigts, B. J. S. Pope, J. Dennis, and P. G. Tuthill, "Differentiable optics with ∂Lux: I—deep calibration of flat field and phase retrieval with automatic differentiation," J. Astron. Telesc. Instruments, Syst. **9** (2023).
4. J. Bradbury, R. Frostig, P. Hawkins, *et al.*, "JAX: composable transformations of Python+NumPy programs," (2018).



\end{document}